\documentclass[%
reprint,
superscriptaddress,
nofootinbib,
amsmath,amssymb,
aps,
]{revtex4-1}

\usepackage{graphicx}
\usepackage{dcolumn}
\usepackage{bm}
\usepackage{hyperref}
\usepackage{color}
\usepackage{dsfont}
\usepackage{pgf,tikz}
\usepackage{pgfplots}
\usepackage{adjustbox}
\usepackage{soul}
\usepackage{rotating} 
\usepackage{enumerate}
\usepackage{subfigure}
\usepackage[normalem]{ulem}

\newcommand{\ket}[1]{|{#1}\rangle}
\newcommand{\ave}[1]{\langle{#1}\rangle}               

\def\Rb87{$^{87}\text{Rb}$}
\def\0{\ket{0}}
\def\1{\ket{1}}

\usepackage{cancel}

\definecolor{mygreen}{rgb}{0,0.5,0}
\definecolor{mygrey}{rgb}{0.5,0.5,0.5}
\definecolor{myred}{rgb}{0.75,0,0}
\definecolor{myblue}{rgb}{0,0,0.75}
\definecolor{mymagenta}{cmyk}{0,1,0,0.12}
\definecolor{mycyan}{cmyk}{1,0,0,0.12}
\definecolor{myorange}{rgb}{1,0.5,0}
\definecolor{myviolet}{rgb}{0.5,0.0,0.75}

\newcommand{\commentout}[1]{}

\begin{document}

\newcommand{\SQL}{{SQL}}
\newcommand{\JPerpSq}{{\hat{J}_{\perp}^2}}
\newcommand{\supmax}{^{\rm max}}
\newcommand{\supmin}{^{\rm min}}
\newcommand{\submax}{_{\rm max}}
\newcommand{\submin}{_{\rm min}}

\newcommand{\ICFO}{ICFO - Institut de Ciencies Fotoniques, The Barcelona Institute of Science and Technology, 08860 Castelldefels, Barcelona, Spain}
\newcommand{\ICREA}{ICREA - Instituci\'{o} Catalana de Recerca i Estudis Avan{\c{c}}ats, 08010 Barcelona, Spain}
\newcommand{\IOPPAS}{Institute of Physics PAS, Aleja Lotnikow 32/46, 02-668 Warszawa, Poland}


\title{Criticality-enhanced quantum sensing in ferromagnetic Bose-Einstein condensates: role of readout measurement and detection noise}
\author{Safoura S. Mirkhalaf}
\affiliation{\IOPPAS}
\author{Daniel Benedicto Orenes}
\affiliation{\ICFO}
\author{Morgan W. Mitchell}
\affiliation{\ICFO}
\affiliation{\ICREA}
\author{Emilia Witkowska}
\affiliation{\IOPPAS}

\date{\today}

\begin{abstract}
We theoretically investigate estimation of the control parameter in a ferromagnetic Bose-Einstein condensate near second order quantum phase transitions.  We quantify sensitivity by quantum and classical Fisher information and using the error-propagation formula. For these different metrics, we find the same, beyond-standard-quantum-limit (SQL) scaling with atom number near critical points, and SQL scaling away from critical points. We find that both depletion of the $m_f=0$ Zeeman sub-level and transverse magnetization provide signals of sufficient quality to saturate the sensitivity scaling. To explore the effect of experimental imperfections, we study the scaling  around criticality at nonzero-temperature and with nonzero detection noise. Our results suggest the feasibility of sub-SQL sensing in  ferromagnetic condensates with current experimental capabilities.
\end{abstract}

\maketitle
\section{Introduction}

In quantum sensing and metrology \cite{GiovannettiS2004, PezzeRMP2018}, a classical parameter such as an externally applied field, an energy level separation, or the elapsed time, is estimated from measurements on a quantum system consisting of $N > 1$ particles. The precision of the estimation is closely related to the sensitivity of the quantum state to the parameter.  It is well known, for example, that entangled states of $N$ two-state particles can estimate a single-particle phase $\phi$ with variance $\delta \phi^{-2} = N^2$, known as the Heisenberg limit (HL), while non-entangled states in the same scenario can at best achieve $\delta \phi^{-2} = N$, known as the standard quantum limit (SQL).  This difference is due to the extreme phase sensitivity of entangled states, for example ``NooN'' states, which become self-orthogonal under a phase shift of $\phi = \pi/N$, when a single particle would become self-orthogonal after a phase shift of $\pi$  \cite{lee2002, MitchellN2004}.  Such states are very sensitive to decoherence, and while it is possible to achieve sensitivities beyond the SQL with large $N$ \cite{TsePRL2019}, the scaling with $N$ returns to the SQL scaling in the presence of very small decoherence~\cite{Demkowicz-Dobrzanski:2012, EscherNP2011}. 

A less-studied scenario, of relevance especially to atomic quantum sensing, is when the quantum state being measured is the result of interactions that act at the same time as the parameter to be measured.  This is the scenario for example in a non-linear interferometer, in which the particles interact while experiencing a phase shift \cite{gross2010nonlinear}. Such system can show counter-intuitive behaviours. For example, if the unknown parameter is itself the strength $\kappa$ of a particular $k$-body interaction, estimation without entanglement yields $\delta \kappa^{-2} \propto N^{2k-1}$, which shows better scaling than the HL already with the simplest two-body interactions \cite{Caves2008, Caves2007, NapolitanoN2011}, and without entanglement.  

Arguably the most natural scenario in which one finds high sensitivity to external parameters is in critical systems, e.g. near a quantum phase transition. These are by nature interacting many-body systems, and we may expect to find advantageous scaling even without entanglement \cite{BraunRMP2018, pra78042105, sachdev2011}. In the case of second-order (continuous) phase transition, the sensitivity in the estimation of control parameter $q$ scales as $\delta q^{-2} \propto N^{2/d\nu}$, where $d$ is the spatial dimension and $\nu$ is the critical exponent of the correlation length~\cite{PhysRevB.81.064418, Rams2018,Pezze2019}. It has already been reported that several many-body models are characterized by $2/d\nu >1$~\cite{LMG_2015, Paris2014, Kwok2008, Buonsante_2012,Rams2018,safoura2020}, and therefore advantageous scaling relative to uncorrelated particles. In the case of first-order phase transition, to the best of our knowledge there is no known bound on $\delta q^{-2}$  \cite{safoura2020}.

In this paper, we theoretically estimate the sensitivity of measurement of coupling constant in a~spin-1~Bose-Einstein condensate system with ferromagnetic interaction \cite{Rotor2010,Duan2013, Ueda2013}. To this end, we make use of the underlying continuous phase transitions and apply the approach mentioned above. 
In particular, we concentrate on the zero longitudinal magnetization case when the phase diagram of ground states exhibits two second-order phase transitions, namely between broken-axisymmetry/antiferromagnetic and  longitudinal polar/broken-axisymmetry phases \cite{Duan2013,ueda2012}. In order to quantify the sensitivity, we make use of the quantum and classical Fisher information as well as the error propagation formula. The quantum Fisher information (QFI) is known as a pivotal parameter of quantum metrology which is linked to the ultimate HL. However, it is not always easy  to find the optimal measurement to saturate the upper bound and extract it in practice. This concerns the fact that in order to find the QFI in experiment, one needs to make a full state tomography of the density matrix which is difficult for large systems. In this case, it is more convenient to use the classical Fisher information (CFI) or error propagation formula which are more easily accessible by making measurement on the appropriate signals.  

For our ferromagnetic spin-1 BEC system with finite size $N$, we derive the QFI and find 
the best sensitivity in estimation of order parameter as $N^{-4/3}$ (implying the sub-shot noise limit) around the critical points. However away from the critical regions, the sensitivity scales as $N^{-1}$ referring to the SQL. This is expected since with no criticality, the neighbor states are indistinguishable with respect to the control parameter and the sensitivity decreases to the classical limit. We confirm this result using the quantum perturbation approach for small values of coupling constant. In addition, we numerically calculate the classical Fisher information as well as the error propagation formula for two particular signals:~$(i)$~the number of atoms in the $m_f=0$ Zeeman state and~$(ii)$~transverse magnetization. 
Both quantities can be used as the order parameter of our system, and we prove that they are both optimal choices of measurement observables in the critical system~\cite{Pezze2019}. 
We show that these sensitivity tools lead to the same power law scaling versus $N$ as the quantum Fisher information which confirms the results in Ref. \cite{Pezze2019}. In particular, our  scaling of the sensitivity of coupling constant around criticality  $\sim N^{-4/3}$ gives the same value as for Lipkin-Meshkow-Glick~\cite{Kwok2008}, Dicke~\cite{Liu2009} and bosonic Josephson junction~\cite{Pezze2019} models as well as the antiferromagnetic spin-1 condensates~\cite{safoura2020} around continuous quantum phase transitions. In this sense, our results suggest that the universal behaviour of the QFI belongs to the same class as aforementioned critical systems.

From the experimental point of view, we propose different types of measurements with respect to our signals, namely the population counting of particles (using absorption imaging or fluorescence imaging) and  paramagnetic Faraday rotation \cite{gajdacz2016preparation,bason2018measurement,palacios2018multi}. 
In addition, in order to model the realistic conditions, we include the thermal and Gaussian detection noises in our work. We show that a large enough thermal noise (compared to the quantum gap) suppresses the sensitivity as expected. Moreover, for the case of finite temperature, our results indicate significant decrease of sensitivity by considering either of the signals. In particular, for a finite system of $500$ atoms,
we show that atom number counting procedure would require the detection noise $\leq 6$ atoms to keep the sub-SQL sensitivity. This is a hard task in practice but still accessible with current experimental techniques\cite{singleA,sherson2010single,zhang2012collective,qu2020probing}. On the other hand regarding the Faraday measurement, the detection noise up to the level of $\sigma \simeq 10^3$ would not affect the sensitivity. 
Consequently, our results suggest that the evaluation of the control parameter is possible in a real experiment with sub-SQL sensitivity using the current state of the art capabilities.

The paper is organized as follows. We start with the introduction to the system, model Hamiltonian, and numerical methods in Section~\ref{sec:method}. We review the basis of estimation theory in Section~\ref{sec2}. Next, in Section~\ref{sec4} we show how the theory can be applied to ferromagnetic condensate at zero temperature. Finally, in Section~\ref{sec5} we carefully analyze the effects of non zero temperature and detection noise with the respect to the relevant experimental realization of the method. The summary and conclusion are given in Section~\ref{sec6}.

\section{System, model and numerical method}\label{sec:method}

We consider the spin-1 Bose-Einstein condensate (atoms in the $F=1$ manifold) in the presence of a homogeneous transverse magnetic field $B$. The system is described by the field vector $\hat{\boldsymbol{\Psi}}=[\hat{\Psi}_1,\hat{\Psi}_0,\hat{\Psi}_{-1}]^{T}$ which components describe atoms in the $m_f=0,\pm 1$ Zeeman states. 
We assume the total atoms number to be of the order {up} to few thousands when the generation of spin domains are energetically costly. We use the single mode approximation (SMA) for the system description~\cite{Ueda2013}. In the SMA the external and internal spin degrees of freedom can be decoupled and the components of the field vector transform to $\hat{\Psi}_{m_f}=\phi(\boldsymbol{r})\hat{a}_{m_f}$, where $\hat{a}_{m_f}$ is the bosonic annihilation operator of an atom in the $m_f$-th Zeeman state. In this case, the system  Hamiltonian casts in the following form~\cite{Rotor2010,Duan2013, Ueda2013} 
\begin{eqnarray}\label{H}
\frac{\hat{H}(q)}{c}=\frac{{\rm sign}(c_2)}{2N}\hat{J}^2-q\hat{N}_0,
\end{eqnarray}
which is composed of two terms: 
the first term resulting from the contact interaction between atoms and the second term indicating the effect of a quadratic Zeeman shifts on the energy levels. 
We set the energy unit to ${c}=N|c_2|\int d{\boldsymbol{r}}|\phi(\boldsymbol{r})|^4$, which is proportional to the density $c\propto \rho=N/V$ for homogeneous systems, and the spin-dependent interaction coefficient $c_2$ which is defined in terms of the s-wave scattering lengths \footnote{The explicit form of $c_2$ is given by $c_2=4\pi\hbar^2(a_0 -a_2)/3m$, where $m$ is the mass of each particle and $a_0(a_2)$ is the s-wave scattering length for spin-1 atoms colliding in symmetric channels of total spin $J=0 \, (J=2)$.}. 
In the following, the negative value of $c_2$ is considered which stands for the ferromagnetic interaction~\cite{Ueda20122, Ueda2013}. 
In equation (\ref{H}), the total pseudo-spin operator squared $\hat{J}^2=\hat{J}_x^2+\hat{J}_y^2+\hat{J}_z^2$ can be expressed in terms of annihilation (creation) operators $\hat{a}_{m_f}(\hat{a}^\dagger_{m_f})$ of an atom in the $m_f$-th Zeeman component, namely 
\begin{align}
	\hat{J}_{x} &\ =\ \frac{1}{\sqrt{2}}\left( \hat{a}^{\dag}_{\scriptscriptstyle{-1}}\hat{a}_{\scriptscriptstyle{0}} + \hat{a}^{\dag}_{\scriptscriptstyle{0}}\hat{a}_{\scriptscriptstyle{-1}} + \hat{a}^{\dag}_{\scriptscriptstyle{0}}\hat{a}_{\scriptscriptstyle{+1}} + \hat{a}^{\dag}_{\scriptscriptstyle{+1}}\hat{a}_{\scriptscriptstyle{0}}\right), \\
	\hat{J}_{y} &\ =\ \frac{i}{\sqrt{2}}\left( \hat{a}^{\dag}_{\scriptscriptstyle{-1}}\hat{a}_{\scriptscriptstyle{0}} - \hat{a}^{\dag}_{\scriptscriptstyle{0}}\hat{a}_{\scriptscriptstyle{-1}} + \hat{a}^{\dag}_{\scriptscriptstyle{0}}\hat{a}_{\scriptscriptstyle{+1}} - \hat{a}^{\dag}_{\scriptscriptstyle{+1}}\hat{a}_{\scriptscriptstyle{0}}\right), \\ 
	\hat{J}_{z} &\ =\  \hat{a}^{\dag}_{\scriptscriptstyle{+1}}\hat{a}_{\scriptscriptstyle{+1}} - \hat{a}^{\dag}_{\scriptscriptstyle{-1}}\hat{a}_{\scriptscriptstyle{-1}}.
\end{align}
In addition, $\hat{N}_{m_f}=\hat{a}_{m_f}^{\dagger}\hat{a}_{m_f}$ is the number operator of atoms in the $m_f$-th Zeeman state, $N$ is the total atom number operator being the eigenvalue of $\hat{N}=\sum_{m_f}\hat{N}_{m_f}$ and the coupling constant $q$ is the strength of the quadratic Zeeman energy. In practice, the parameter $q$ can be expressed as a sum of two terms, $q=q_{\rm B} + q_{\rm M}$, as it can be tuned using an applied magnetic field or an off-resonant microwave dressing field \cite{dressing2014,dressing2015}. Therefore, the value of $q$ can be controlled experimentally from negative to positive values~\cite{dressing2014,dressing2015}. 
The model can be realized with ultra-cold $^{87}$Rb atoms populating the three magnetic sublevels of the $F=1$ ground state manifold~\cite{Stamper_Kurn_1998,PhysRevLett.87.010404}.

The $z$ component of pseudo-spin $\hat{J}_z$ is a constant of motion, as $[\hat{H}, \hat{J}_z]=0$.
The eigenvalues of $\hat{J}_z$ correspond to the longitudinal magnetization of the system $M = N_{+1} - N_{-1}$ which is conserved and, therefore, can be used to label the Hamiltonian eigenstates~\footnote{It is worth to note that $\hat{J}^2$ is also an additional conserved quantity and its eigenvalues, in addition to $M$, can also be used to label the Hamiltonian eigenstates. We use this fact for the analytical description of sensitivity around $q=0$, see in Appendix~\ref{app:perturbation}.}. In this case, the system Hamiltonian has the block diagonal structure while eigenstates can be considered in the subspace of fixed magnetization. This is also justified based on the fact that the spin-dependent interaction has rotational symmetry as long as the spin-1 system is isolated from its environment and dipolar interactions are neglected~\cite{Gerbier2012}. Therefore, the linear Zeeman energy acts as a constant shift on the energy levels. 
In this paper, we consider even values of the total number of atoms $N$ and zero longitudinal magnetization $\ave{\hat{J}_z}\equiv M=0$. 

In order to describe numerically the system Hamiltonian and extract its eigenstates, we employ the Fock basis constituted by all eigenstates of the atomic number operators $\hat{N}_{m_f}$.
We use the parametrization
$|n\rangle=|N_{1},N_0,N_{-1}\rangle=|n,N+M-2n,n-M\rangle $,
with $n\in [n_{\rm min}, n_{\rm max}]$ where $n_{\rm min}={\rm max}[0,M/2,M]$ and $n_{\rm max}={\rm min}[N,(M+N)/2,N+M]$.
Subsequently, we build up the Hamiltonian 
(\ref{H}) in this basis and numerically diagonalize it to  obtain the respective eigenenergies $E_{\alpha}$ and eigenstates $|\psi_{\alpha}\rangle$. 
The ground state (GS) $|\psi_0\rangle$ is used to estimate the sensitivity at zero temperature. 
When the temperature value is non-zero, the quantum state is represented by the canonical Gibbs density matrix 
\begin{eqnarray}
\label{rho}
\hat{\rho}(q,T)=\sum_{\alpha} \frac{e^{-E_{\alpha}(q)/k_BT}}{Z} |\psi_\alpha\rangle \langle \psi_{\alpha}|,
\end{eqnarray}
where the $\alpha$-th eigenstate is weighted by $w_{\alpha}={e^{-E_{\alpha}(q)/k_BT}}/{Z}$ and $Z=\sum_n e^{-E_\alpha/k_B T}$ is the partition function with the Boltzamann constant $k_B$. Note that in the zero temperature limit, the quantum state of the system approaches GS while in the high temperature limit the quantum state is maximally mixed with equally populated eigenstates, i.e. they have the same weight $w_{\alpha}$. 

\newcommand{\AFM}{AFM}

\begin{figure}[]
	\includegraphics[width=1.\columnwidth]{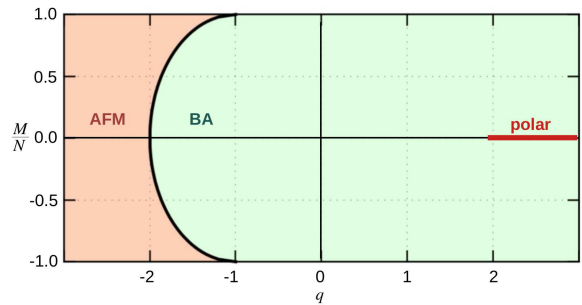}
	\caption{The mean-field phase diagram of the ground states of the ferromagnetic spin-1 Bose-Einstein condensate~\cite{Ueda20122, Ueda2013}. 
	\emph{The \AFM~phase:} atoms coexist in the $m_F=\pm 1$ Zeeman components. 
	\emph{The BA phase:} atoms occupy all three Zeeman components. 
	\emph{The polar phase:} all atoms are in the $m_F=0$ Zeeman component. 
	In general, the ground state is a superposition of Fock states. 
	However, when $q\ll - 2$ the ground state of AFM phase simplifies as $|(N+M)/2,0,(N-M)/2\rangle$ while for $q\gg 2$ the ground state of the polar phase is $|0,N,0\rangle$. 
	The position of the left critical point within the mean-field description is given by the formula $q_c=-1-\sqrt{1-(M/N)^2}$.}
	\label{fig:fig1}
\end{figure}

The phase diagram of ground states of the Hamiltonian~(\ref{H}) is presented in Fig.~\ref{fig:fig1}. 
The three phases can be distinguished: the longitudinal polar, broken-axisymmetry (BA) and antiferromagnetic (\AFM) one \cite{Duan2013,Ueda20122}. 
In particular, for the zero magnetization case, $M=0$, two phase transitions occur between BA/\AFM~ and longitudinal polar/BA phases at the values of control parameter $q_c=-2$ (the left critical point) and $q_c=2$ (the right critical point), respectively. These two critical points are our central interest in this work.

In the next section, we give a brief review of the basics of the quantum estimation theory which we use to estimate the sensitivity of measurement of $q$ around criticality.

\section{Quantum estimation theory}\label{sec2}

When a physical system crosses a critical point by varying a control parameter, the ground state of the system changes abruptly, indicating the presence of a quantum phase transition. A direct consequence of this fact is significant distinguishability of the lowest energy eigenstates of the system around criticality. 
It has been reported that this distinghisability can be used to enhance the precision in estimation of the control parameter around criticality~\cite{Zanardi2007,paris2009,Paris2016,Ram2018}. 
In the following we recall the building-blocks of the estimation theory used in the paper.

Let us start with the generic Hamiltonian
\begin{equation}\label{eq:Hq}
\hat{H}=\hat{H}_0+q\hat{H}_q,
\end{equation}
for which a corresponding state $\hat{\rho}$ exhibits a quantum phase transition around the critical value of $q=q_c$. This means that a change of the control parameter $q$ around $q_c$ by infinitesimally small amount $\delta_q$ leads to a magnificent change between the respective ground states, say $\hat{\rho}({q})$ and $\hat{\rho}(q+\delta_q)$. The amount of distiguishibility between these two states is quantified by the fidelity \cite{UHLMANN1976273}, 
\begin{equation}\label{fidel}
\mathcal{F}_q={\rm tr}\left[\sqrt{\hat{\rho}(q)}\hat{\rho}(q+\delta_q)\sqrt{\hat{\rho}(q)}\right]^{1/2}.
\end{equation}
In the special case of pure state $|\psi(q)\rangle$, the fidelity simplifies to $\mathcal{F}_q=|\ave{\psi(q)|\psi(q+\delta_q)}|$.  

In the estimation theory, the notion of fidelity is used to evaluate the best precision in estimation of the $q$ parameter and is set by the quantum Fisher information~\cite{caves1994,Filho2013,You2007,You2015}
\begin{equation}\label{QFI_fid}
F_q=-4 \, \frac{\partial^2 \mathcal{F}_q}{\partial \delta_q{}^2} \, \left|_{\delta_q\to 0} \right. .
\end{equation}
More precisely, the precision is determined by the inverse of the QFI, i.e. the larger the value of the QFI, the better the precision.
The QFI value increases around criticality because of the abrupt change of ground states of the system around the critical points. Therefore, the QFI can be used as a criterion for distinguishability of the quantum states around phase transitions. 
Subsequently, the pivotal role of QFI in the  theory is precise evaluation of the best possible sensitivity for evaluation of $q$~\cite{caves1994, Pezze:2016_review}. 
The QFI determines the upper bound on the sensitivity in the parameter estimation which is the quantum Cramer-Rao bound (QCRB) \cite{caves1994}. 
The usage of quantum resources might overcome the ultimate limit for the precision reachable with the uncorrelated particles, namely the so-called SQL. 

In practice, the measurement of the QFI is a complex task and it requires a full quantum state tomography which is not very feasible for large systems using the current experimental techniques. 
In this case, it is more convenient to consider the classical Fisher information introduced based on the probability distributions of being $\hat{\rho}$ in eigenstates of observable $\hat{\mathcal{S}}$, namely  $P(s|q)=\ave{s|\hat{\rho}(q)|s}$ (here, $|s\rangle$ and $s$ are eigenstates and eigenvalues of $\hat{\mathcal{S}}$, respectively). 
The CFI is defined as
\begin{eqnarray}\label{clFI_fid}
F_{c}=-4\, \frac{\partial^2 \mathcal{F}_{c}}{\partial \delta_q^2},
\end{eqnarray}
where the fidelity between two neighbor probability distributions in statistical space is
\begin{equation}\label{eq:classfidelity}
\mathcal{F}_c=\sum_{s} \sqrt{P(s|q)P(s|q + \delta_q)} .
\end{equation}
In order to evaluate the probability distribution $P(s|q)$, one requires the knowledge of the whole basis of $\hat{\mathcal{S}}$. In real experiments, the CFI can be indirectly evaluated based on measurements of the Hellinger distance between probability distributions of the measured observable $\hat{\mathcal{S}}$ for neighbouring states $\hat{\rho}(q)$ and $\hat{\rho}(q+\delta q)$ ~\cite{Strobel:2014, PhysRevX.10.011018}.
The QFI is defined as the maximization of the CFI over all possible $\hat{\mathcal{S}}$~\cite{Pezze:2016_review}. In the case of single parameter estimation, it is proved that there is always one measurement basis which saturated the QCRB. However, since the {optimal measurement basis} is not always evident, choosing the best one is a non-trivial task on its own \cite{Pezze:2016_review}. 
In addition, the precision in the estimation of an unknown parameter $q$ can also be evaluated using the standard error-propagation formula
\begin{equation}
\delta q^2
=\frac{\Delta^2{\hat{\mathcal{S}}}}{|\partial_q\ave{\hat{\mathcal{S}}}|^2} ,
\label{delta}
\end{equation}
with $\Delta^2{\hat{\mathcal{S}}}=\ave{\hat{\mathcal{S}}^2}-\ave{\hat{\mathcal{S}}}^2$ representing the variance of $\hat{\mathcal{S}}$. 

\begin{figure}[hbt!]
	\includegraphics[width=1.\columnwidth]{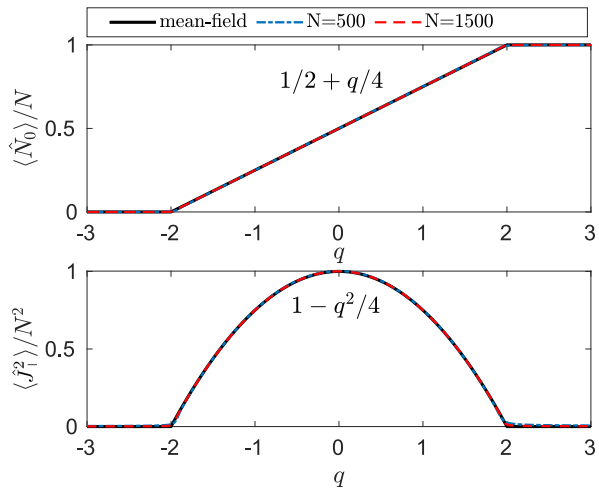}
	\caption{An average value of the fraction of atoms $n_0$ in the $m_F=0$ Zeeman component (upper panel) and the transverse magnetization squared $j^2$ (lower panel) for $M=0$. The black solid lines show analytical results, (\ref{mf_sig_n0}) and (\ref{mf_sig_j2}), obtained in the mean-field limit. The dashed blue and dash-dotted red lines represent numerical results obtained using the exact diagonalization method for $N=500$ and $N=1500$, respectively.}
	\label{fig:fig2}
\end{figure}

In this work, we consider the two different observables: $(i)$ the atomic population in the $m_F=0$ Zeeman component $\hat{\mathcal{S}}=\hat{N}_0$, and $(ii)$ the transverse magnetization $\hat{J}^2_\perp=\hat{J}^2_x+\hat{J}^2_y$.
An average value of both the observables,
\begin{equation}
n_0=\frac{\ave{\hat{N}_0}}{N}, \ \ \ {j}=\frac{\sqrt{\ave{\hat{J}^2_\perp}}}{N}\equiv \frac{\sqrt{\ave{\hat{J}^2}}}{N},
\end{equation}
have already been recognized as the order parameters to characterize the respective two phase  transitions at $q=\pm 2$~\cite{Zurek2007,Lama2007,Chapman2016,Ming2018}. 
Note the equivalence of ${\ave{\hat{J}^2_\perp}}$ with ${\ave{\hat{J}^2}}$ which arises due to the fact that we are considering $\ave{\hat{J}_z}=0$ and the assumption of fixed magnetization~\footnote{By fixed magnetization, we mean $\Delta^2 \hat{J}_z=0.$}.
In Fig. \ref{fig:fig2} we show the change of both order parameters with respect to the coupling constant $q$. The first order parameter $n_0$, can be measured experimentally by measuring the atomic population in the $m_f=0$ Zeeman state. The transverse magnetization ${j}$ can be obtained using paramagnetic Faraday rotation measurements (we discuss the details in Section \ref{sec5}).

In general, the imprecision $\delta q$ satisfies the inequalities
\begin{equation}\label{eq:inequalities}
    \delta q^{-2} \le {F_{c}} \le {F_q}.
\end{equation}
As mentioned before, the QFI gives the highest possible sensitivity to $q$ (QCRB) at the expense of experimental difficulties of state tomography. On the other hand, the error-propagation formula gives the lowest sensitivity while it needs only measurement of the first and second moments of the observable $\hat{\mathcal{S}}$ which is a bonus from the experimental point of view. Meanwhile, evaluating the CFI relies on the extracting the higher moments of $\mathcal{S}$ which leads to higher sensitivity than the signal-to-noise ratio evaluation. 
In the following, we discuss the enhancement in the estimation of the control parameter $q$ using the quantum estimation theory around criticality.

\section{precise estimation of coupling constant around critical points} \label{sec4}
\begin{figure*}[hbt!]
	\begin{picture}(0,160)
	\put(-265,0){\includegraphics[width=.37\textwidth]{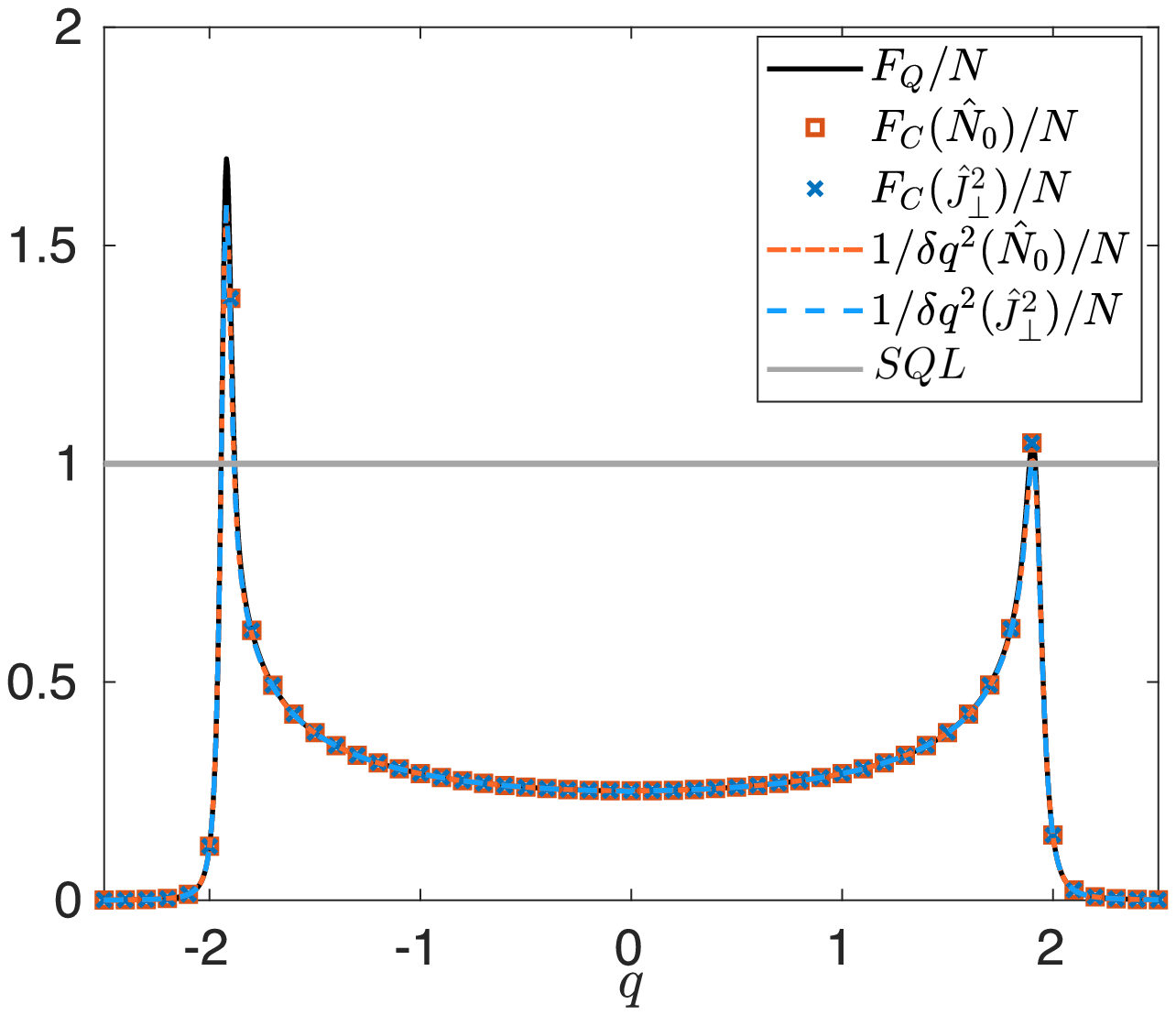}}
	\put(-236,126){(a)}
	\put(-93,0){\includegraphics[width=.37\textwidth]{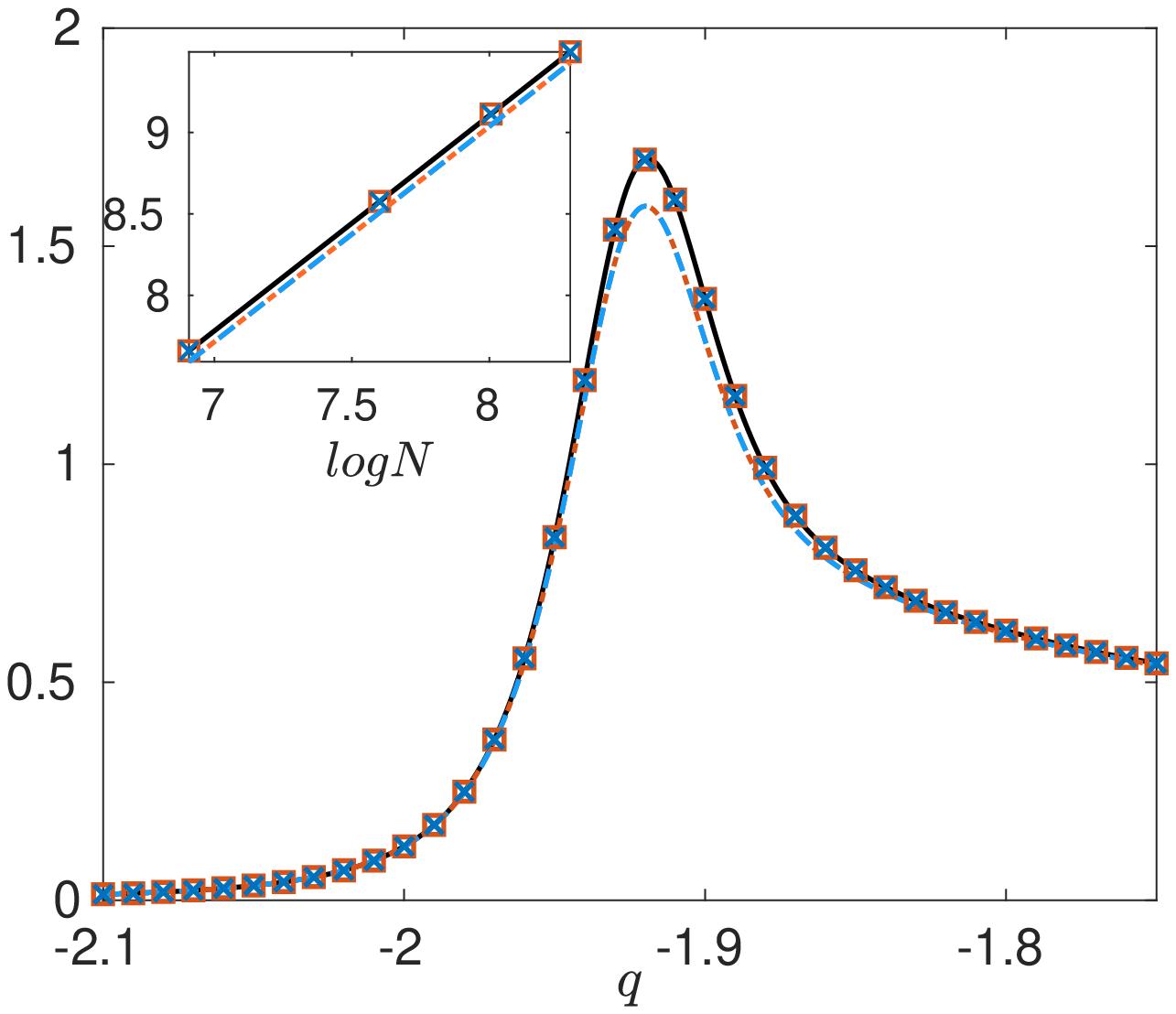}}
	\put(60,126){(b)}
	\put(80,0){\includegraphics[width=.37\textwidth]{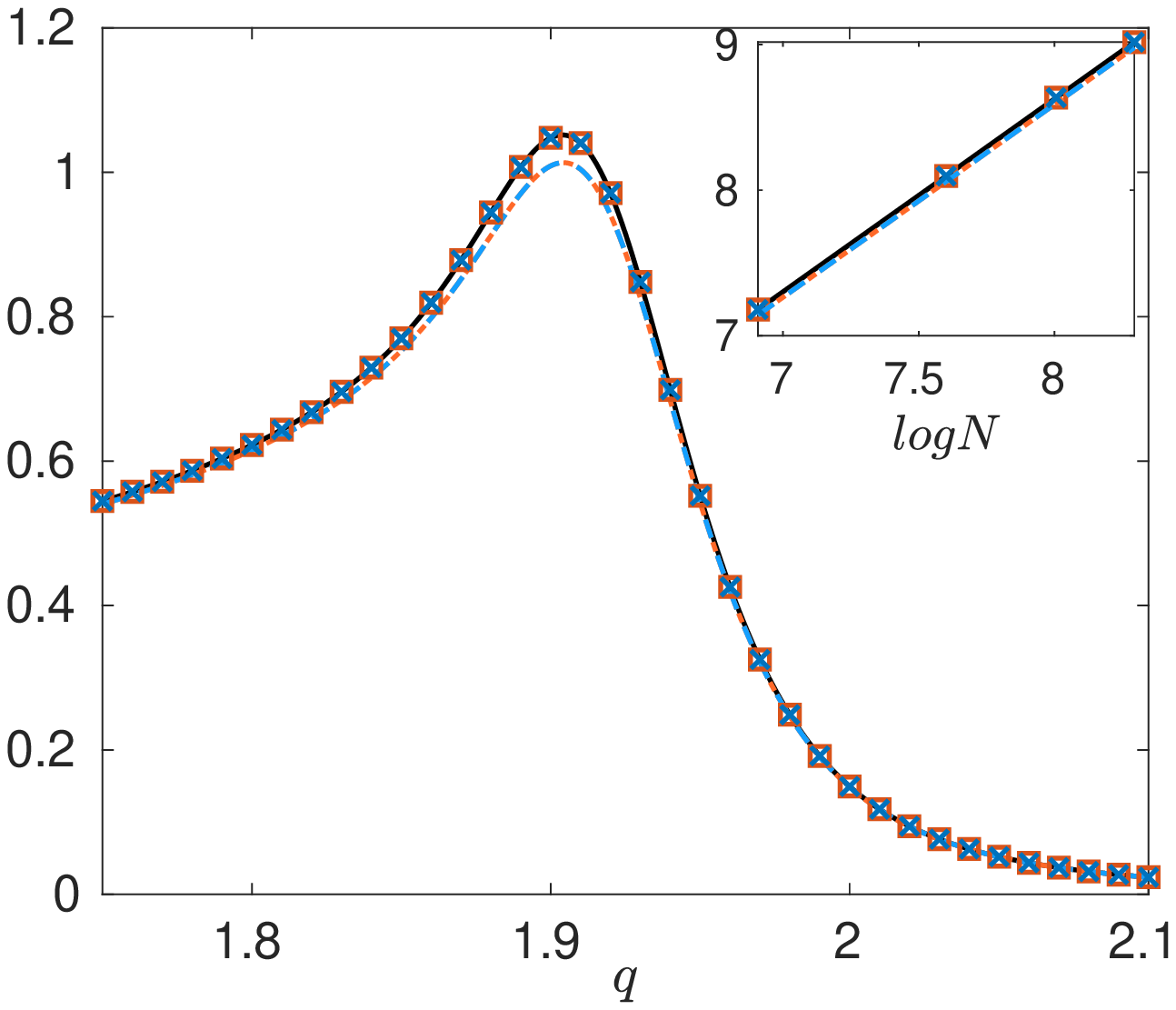}}
	\put(110,126){(c)}
	\end{picture}
	\caption{
		(a) Variations of the QFI (black solid line), CFI with $\hat{\mathcal{S}}=\hat{N}_0$ (red squares) and $\hat{\mathcal{S}}=\JPerpSq$ (blue crosses), and $\delta q^{-2}$ with $\hat{\mathcal{S}}=\hat{N}_0$ (the dot-dashed red line) and $\hat{\mathcal{S}}=\hat{J}_{\perp}^2$ (the turquoise dashed line) for the total number of atoms $N=500$, $M=0$ and zero temperature. The SQL is marked by the thick solid gray line. All quantities are divided by $N$.
		The panels (b) and (c) show the same but for the values of $q$ closely around the left and right critical points, respectively. 
		A scaling of the best sensitivity (quantified by the maxima of QFI, CFI and $\delta^{-2}q$) versus the total atoms number is presented in insets of (b) and (c). Note, we plot logarithms of these quantities to demonstrate the power law dependence.
		Specifically, we observe that $F_q\supmax=F_c\supmax\propto 0.22N^{1.33}$ and $\delta q^{-2,\rm max}\propto 0.21N^{1.33}$ around both critical points.}
	\label{fig:fig3}
\end{figure*}

In this section, we employ the QFI, CFI, and error-propagation formula to discuss characteristic features of precision in the estimation of $q$ around two critical regions for zero temperature. In Fig. \ref{fig:fig3} we present the numerical value of these quantities for $\hat{\mathcal{S}}=\hat{N}_0$ and $\hat{\mathcal{S}}=\hat{J}_\perp^2$ as a function of $q$ for $N=500$. The significant increase of all three quantities at the two critical points around $q=\pm 2$ is observed. The appearance of peaks reflects the significant distinguishability of the ground states around each critical point. The heights of peaks are different because ground states on both sides of critical points have various character: BA/AFM for the left and polar/BA for the right crtitical points.

The almost ideal overlap among all curves can be noticed in Fig.~\ref{fig:fig2}, and it demonstrates that the CFI saturates the QCRB while $\delta q^{-2}$ is slightly reduced by the same amount for both signals. 
Still, results presented in Fig.~\ref{fig:fig3} show that the measurement of both order parameters leads to the estimation of $q$ with sub-SQL sensitivity around the critical points (more details in the following). 
Interestingly, both signals give the same value for the CFI and $\delta q^{-2}$. 
Specifically, for the error propagation formula
\begin{equation}\label{err}
\frac{\Delta^2 \hat{N}_0}{|\partial_q\ave{{\hat{N}_0}}|^2}
\approx\frac{\Delta^2 \JPerpSq}{|\partial_q\ave{{\JPerpSq}}|^2},
\end{equation}
can be explained analytically. This is shown in Appendix~\ref{app:explanation} using the fact that variation of the Hamiltonian (\ref{H}) tends to zero and estimating the first and second moments of $\hat{\mathcal{S}}$ by using mean-field approximation. 
Although theoretically, the choice of order parameter seems neutral for both $\hat{N}_0$ and $\JPerpSq$, the respective measurements are inherently different from an experimental point of view. This has implications when considering the effect of the noise in the detection process. This point will be discussed in Section \ref{sec5}.

On the other hand, it is expected that the maximum value of the Fisher information with respect to number of particles $N$  is subject to a power-law scaling \cite{Zanardi2007,Rams2018,Pezze2019}. 
\begin{equation}
F_q\supmax \propto N^{\mu}
\end{equation}
with $\mu=2/d\nu$ where $\nu$ is the critical exponent describing the divergence of correlation length and $d$ is the effective spatial dimension, as explained in \cite{Rams2018,Pezze2019}.
In order to demonstrate how the sensitivity changes by varying the total number of atoms, in insets of Figs.~\ref{fig:fig3}(b) and~\ref{fig:fig3}(c), we show the logarithmic values for  $F_q\supmax,\, F_c\supmax$ and $(\delta q^{-2})\supmax$ versus $\log N$. 
Indeed, we observe the power law scaling of the QFI with $\mu=4/3$. The same scaling exponent for the QFI (or equivalently fidelity susceptibility) in the Lipkin-Meshkov-Glick~\cite{Kwok2008}, Dicke~\cite{Liu2009}, bosonic Josephson junction~\cite{Pezze2019}, or antiferromagnetic spinor condensate \cite{PhysRevA.101.043609} models. Moreover, we have extracted the same scaling law for the other estimating tools in vicinity of the criticality , say $F_q\supmax\sim F_c\supmax \sim (\delta q^{-2})\supmax \sim N^{4/3}$. This result confirms the findings in \cite{Pezze2019}, that the scaling of the maxima for CFI and signal-to-noise ratio with $N$ coincides with the scaling for QFI provided that the signal is chosen as the order parameter of corresponding continuous phase transition. In this sense, the order parameter gives the optimal measurement basis for evaluating CFI or error-propagation formula.

We would like to stress that the scaling of the sensitivity around critical points with either of these metrological tools beats the scaling of SQL. On the other hand, we can show that the sensitivity is of the order of SQL in between the two critical points. In particular, using perturbation theory around $q=0$, we can prove that $F_q\sim F_c \sim \delta q^{-2} = N/4$ which is in agreement with the numerical predictions presented in Fig.~\ref{fig:fig3} (the details of analytical calculations are discussed in Appendix~\ref{app:perturbation}). Consequently, by tuning the coupling constant from the positive to negative values \cite{dressing2014,dressing2015}, the sensitivity of estimating $q$ around the critical point is enhanced. 

Up to now, we have considered the theoretical $q$-estimation protocol under ideal conditions, that is the zero temperature regime and perfect measurements of the signals. In the following, we will carefully address the effect of detection noise and finite temperature. These are real experimental constrains that can reduce the signal-to-noise ratio and thus reduce the sensitivity.

\section{Experimental protocol and sources of noise}\label{sec5}

The experimental protocol to measure ${\cal \hat{S}} \in \{\hat{N}_0, \JPerpSq \}$, and estimate $q$ in the spinor BEC would follow the steps: $(i)$ \textit{State preparation}. A sample of atoms is cooled down through forced evaporation to reach the BEC phase in its polar state i.e $\ket{0,N,0}$ and $q(t=0) > 2$; $(ii)$ \textit{Adiabatic evolution}. An external field such as microwave dressing field is used to adiabatically change the control parameter to a final value $q(t_f)$. This change needs to be slow enough to fulfill the adiabaticity condition $\Delta \hat{H} \ll {\hbar}/{\tau}$, and specifically at the quantum phase transition $\Delta t \ll {\hbar}/{\Delta\submin}$ \cite{dressing2014,dressing2015}; $(iii)$ \textit{Detection}. Perform a measurement of ${\cal \hat{S}}$, to find the specific realization ${\cal S}_i$.   $(iv)$ \textit{Estimation}.  Apply a suitable estimator, e.g. the maximum-likelihood estimator, to find the estimate $q_i = \hat{q}({\cal S}_i)$.  The measurement can be repeated to obtain a collection of estimates $\{ q_i \}$, from which the variance $\delta q^2$ can be estimated.  A similar protocol has  been used to demonstrate sub-shot noise sensitivity in interferometric measurements \cite{zou2018beating}.

This experimental protocol, even if performed with extreme care and by a skilled experimentalist, will suffer from several limitations. 
First, no real experiment works in the zero temperature limit. Second, the inevitable detection noise can also diminish the sensitivity of the measurements. Thus, it is compelling to address the effects of these constrains theoretically. 

Although we have considered the average value of both signals $\hat{\mathcal{S}}=\hat{N}_0, \, \JPerpSq$ as the order parameters of the system, there are fundamental differences between the two. In fact, taking $\hat{N}_0$ as our observable leads to a detection process based on the population counting of particles, typically done in ultra-cold quantum gases experiments using absorption imaging or fluorescence imaging. On the other hand, measurements of $\JPerpSq$ can be performed using paramagnetic Faraday rotation. 

In the following, we show how the finite temperature and detection noise affect the measurement of both observables and discuss which detection method could be more resilient against these sources of noise.

\subsection{Effect of non-zero temperature}

\begin{figure}
	\begin{picture}(0,245)
	\put(-135,0){\includegraphics[width=0.53\textwidth]{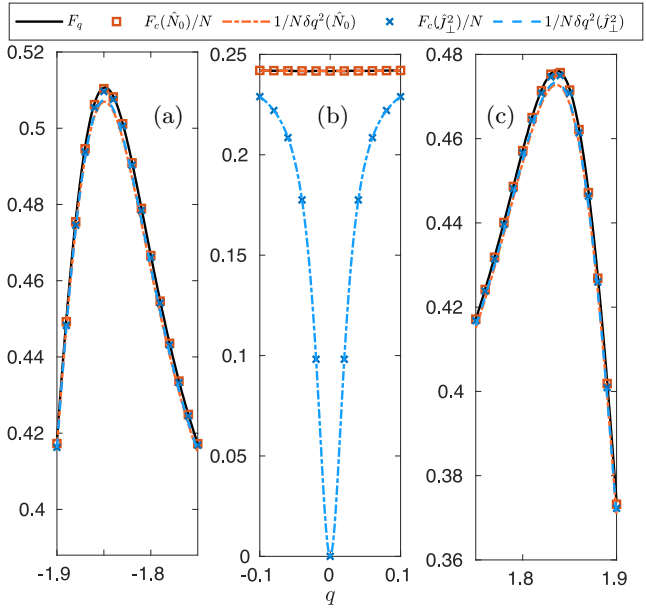}}
	\put(-65,192){(a)}
	\put(-3,192){(b)}
	\put(61,192){(c)}
	\end{picture}
	\caption{The sensitivity quantified by the QFI, CFI and $\delta q^{-2}$ at non-zero temperature $k_BT/c=0.5$, with both  $\hat{\mathcal{S}}=\hat{N}_0, \JPerpSq$, around the left critical point (a), $q=0$ (b) and the right critical point (c). All quantities are divided by $N$. Note the appearance of a dip for the CFI and $\delta q^{-2}$ when $\hat{\mathcal{S}}=\JPerpSq$ around $q\sim 0$ as compared to zero temperature case, Fig.~\ref{fig:fig3}. We observe a shift of critical points (defined as maxima of the QFI) due to non-zero temperature.}
	\label{fig:fig4a}
\end{figure}

The effect of finite temperature is assessed using the density matrix formalism within the canonical Gibbs ensemble (\ref{rho}). 
The overall behavior of $F_q,F_c$ and $\delta q^{-2}$ as a function of the control parameter $q$ for $k_BT/c=0.5$ is presented in Fig.~\ref{fig:fig4a}. 
In general, we observe that the non-zero temperature introduces two characteristic features:~$(i)$~reduction of the QFI, CFI, $\delta q^{-2}$ values and~$(ii)$~appearance of a dip for the CFI and $\delta q^{-2}$ when $\hat{\mathcal{S}}=\JPerpSq$ around $q\sim 0$ as compared to zero temperature case shown in Fig.~\ref{fig:fig3} (both for $N=500$).
Since we are mostly interested in the estimation of $q$ with sub-SQL sensitivity which takes place around critical points, we skip more details about the properties of the dip at $q\sim 0$ in the main text.
Its origin as well as characteristics at non-zero temperature is explained in Appendix \ref{app:finiteT} using perturbation theory.

\begin{figure}
	\begin{picture}(0,418)
	\put(-120,205){\includegraphics[width=.5\textwidth]{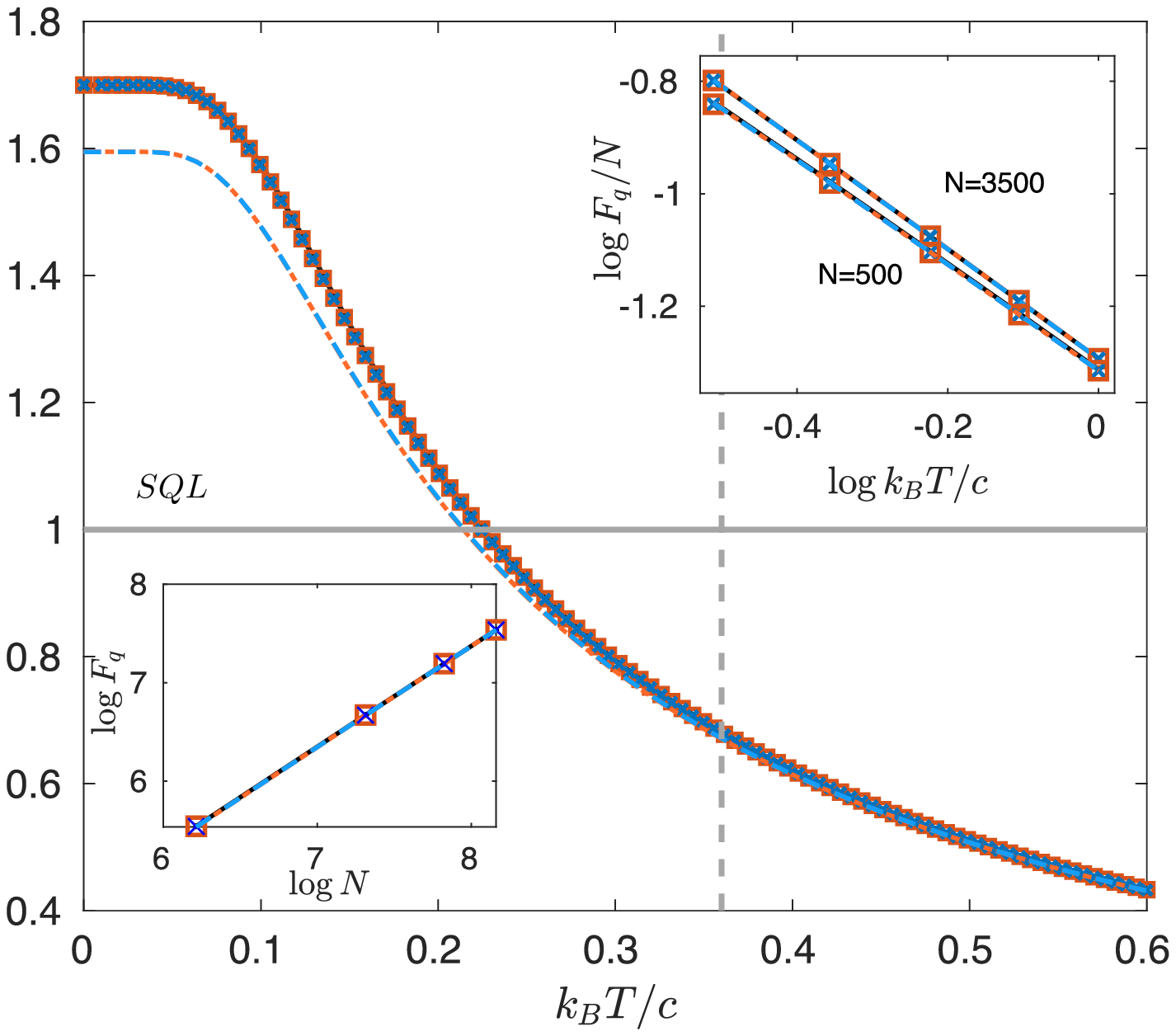}}
	\put(85,260){(a)}
	\put(-120,-8){\includegraphics[width=.5\textwidth]{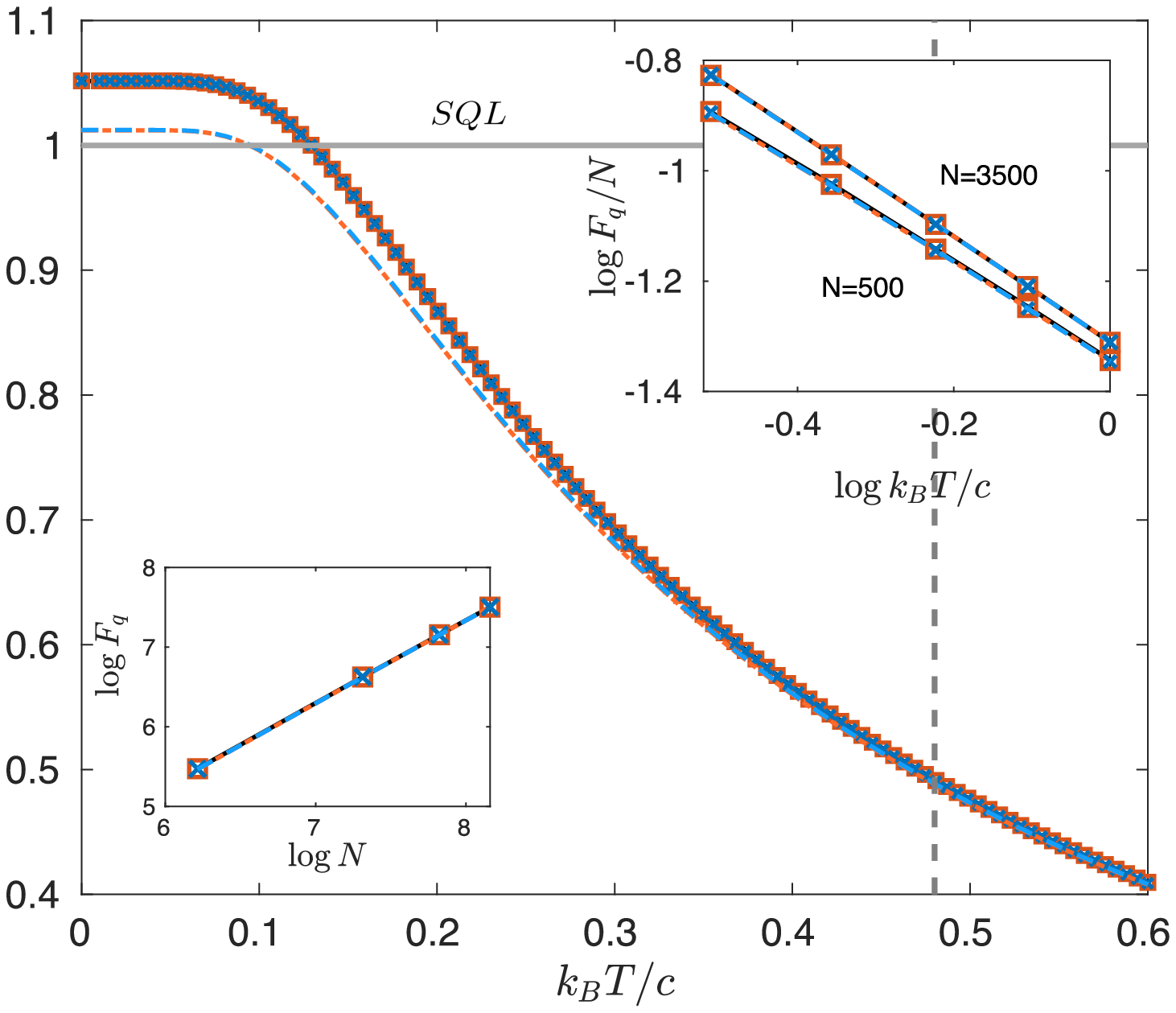}}
	\put(85,40){(b)}
	\end{picture}
	\caption{
		Effect of non-zero temperature on the best sensitivity around the left (a) and right (b) critical points.
		The values of the maxima of the QFI, CFI and $\delta q^{-2}$ with $\hat{\mathcal{S}}=\hat{N}_0,\, \JPerpSq$ versus temperature $k_BT/c$ for $N=500$ and $M/N=0$. 
		All quantities are divided by $N$.
		The same markings as in Fig. 4 is used.
		We have removed the superscript \textit{max} for the case of simplicity.
		The SQL is marked by the horizontal solid gray line. The values of energy gap $\Delta_{\rm min}/c=0.36$ (a) and $\Delta_{\rm min}/c=0.48$ (b) are marked by the vertical dashed lines. 
		The upper right insets demonstrate scaling of the best sensitivity versus $T$ in the high temperature limit, when $k_BT \gg \Delta_{\rm min}$. 
		The power law scaling $T^{\eta}$ is expected  \cite{Ram2018,Ming2018}.
		We observe $\eta\sim 0.93$ for $N=500$, $\eta\sim 0.95$ for $N=1500$ and $\eta\sim 0.97$ for $N=3500$ at the left critical point (a).
		Similarly, we obtained $\eta\sim 0.88 \, (N=500), \, 0.92 \, (N=1500)$ and $0.95 \, (N=3500)$ at the right critical point (b).
	    {The difference between the scaling of left and right peaks seems to be  consequence of the depth of high temperature limit with respect to the respective gaps.} 
		The lower left insets gives the scaling with the total atoms number $N$ in the classical high temperature limit for $kT/c =0.5$. We observe the scaling approaches SQL, $F_q\sim N^{1.02}$, as for uncorrelated particles.		
		}
	\label{fig:fig4}
\end{figure}

In Fig. \ref{fig:fig4} we show the temperature dependence of the maxima of the QFI, CFI and $\delta q^{-2}$ around the left (a) and right (b) critical points. 
Clearly, the finite temperature diminishes the sensitivity for either of the signals due to the fact that the pure ground state transfers to the classical mixture~(\ref{rho}). 
The three different regimes can be distinguished~\cite{safoura2020,Rams2018} depending on the ratio between temperature and the energy gap between the ground and first excited state $\Delta_{\rm min}$: $(i)$ the quantum regime for $k_BT/c \ll \Delta_{\rm min}$, $(ii)$ the intermediate when $k_BT/c\approx \Delta_{\rm min}$ and $(iii)$ the classical one for $k_BT/c\gg\Delta_{\rm min}$. In the quantum regime, the QFI, CFI and $\delta q^{-2}$ are robust against thermal fluctuations while by increasing the temperature, the QFI, CFI and $\delta q^{-2}$ decrease. 
We observe that the rate of change with temperature is similar at both critical points.

In the high temperature limit, all the three quantities show similar behaviour.
This is illustrated in the left upper insets where the logarithms of maxima of $F_q,F_c$ and $\delta q^{-2}$ versus logarithm of temperature are plotted for different values of the total atoms number $N=500, 1500$ and $3500$. 
In fact, the maxima are supposed to be subject to the scaling law $F_q\supmax/N\sim F_c\supmax/N\sim (\delta q^{-2})\supmax/N \sim T^{\eta}$, with $\eta={(d \nu-2)/(z \nu)}$, in terms of $\nu$ and $z$ as the correlation and dynamical critical exponents and $d$ as the effective spatial dimension~\cite{Ram2018}.
In the case of ferromagnetic system, one has $d \nu=3/2$ and $z \nu=1/2$~\cite{Ming2018,Chapman2016}, which leads to $T^{-1}$ as demonstrated in the insets of Fig.~\ref{fig:fig4}. 
{The maxima of $F_c$ and $\delta q^{-2}$ are insignificantly smaller than $F_q$ (not noticeable in insets of Fig.~\ref{fig:fig4}), and are subject to the same scaling laws but with a slightly smaller pre-factors.}
It is worth to note that antiferromagnetic spinor condensates exhibit the same scaling laws as reported in~\cite{safoura2020}. 

On the other hand to investigate the classical high temperature limits, in Fig. \ref{fig:fig4} we have present the logarithmic values of $F_q\supmax,F_c\supmax$ and $(\delta q^{-2})\supmax$ versus $\log N$ in the bottom left insets for $k_B T/c=0.5$ and $N=500$. As compared to the quantum limit shown in the subsets of Figs. \ref{fig:fig3}, our numerical results confirm that in the classical limit, the finite-size scaling of sensitivity decreases from $N^{4/3}$ (sub-SQL) to $N$ (SQL).

\subsection{Effect of detection noise} 

In addition to the temperature, we also consider the effect of detection noise $\sigma$. This noise can have several origins. Here we will assume that it is strictly related to the imperfection detection process, and we do not take into account the shot to shot noise in typical experimental repetitions. In order to include the effect in our theory, we consider the Gaussian \emph{blurring} of the probability distribution $P(s|q)=\ave{s|\hat{\rho}|s}$ as \cite{Pezze:2013,Pezze:2016_review}

\begin{equation}
\label{prob}
\tilde{P}(s|q)=\frac{1}{\mathcal{N}} \sum_{s'} e^{-\frac{(s-{s}')^2}{2\sigma^2}}  P(s'|q)
\end{equation}
with $\mathcal{N}=\sum_{s}{e^{\frac{(s-s')^2}{2\sigma^2}}}$ being the normalization factor~\footnote{This arises due to the general normalization condition over probability distributions $\sum_{s}P_s=1$.}. In order to include Gaussian detection noise in the CFI, one has to replace the probability distribution $P(s|q)$ with $\tilde{P}(s|q)$ in the fidelity (\ref{eq:classfidelity}). Moreover, in the error propagation formula~(\ref{delta}), we change $\ave{\hat{\mathcal{S}}^2}$ and $\ave{\mathcal{\hat{S}}}$ making use of 
(\ref{prob}). It means that the $k$th moment of $\hat{\mathcal{S}}$ under the detection noise reads
\begin{equation}\label{eq:momentsdetectionnoise}
    \langle \hat{\mathcal{S}}^k \rangle_{\rm dn} = \sum_s s^k \tilde{P}(s|q).
\end{equation}
Note that the effect of detection noise on moments of the operator $\hat{\mathcal{S}}$ in thermodynamic limit is the same as if it was replaced by $\hat{\tilde{\mathcal{S}}}=\hat{\mathcal{S}} + \hat{\delta}_\mathcal{S}$, where $\hat{\delta}_\mathcal{S}$ is an independent Gaussian operator satisfying $\ave{\hat{\delta}_\mathcal{S}^{2k+1}}=0$ and $\ave{\hat{\delta}_\mathcal{S}^{2k}}=\sigma^{2k}(2k-1)!!$~\cite{PhysRevA.73.013814}.
However in our calculations, we employ (\ref{eq:momentsdetectionnoise}) since the ensembles is of finite size and not necessarily in the thermodynamic limit. 

\begin{figure}[hbt!]
	\begin{picture}(0,405)
	\put(-130,205){\includegraphics[width=.52\textwidth]{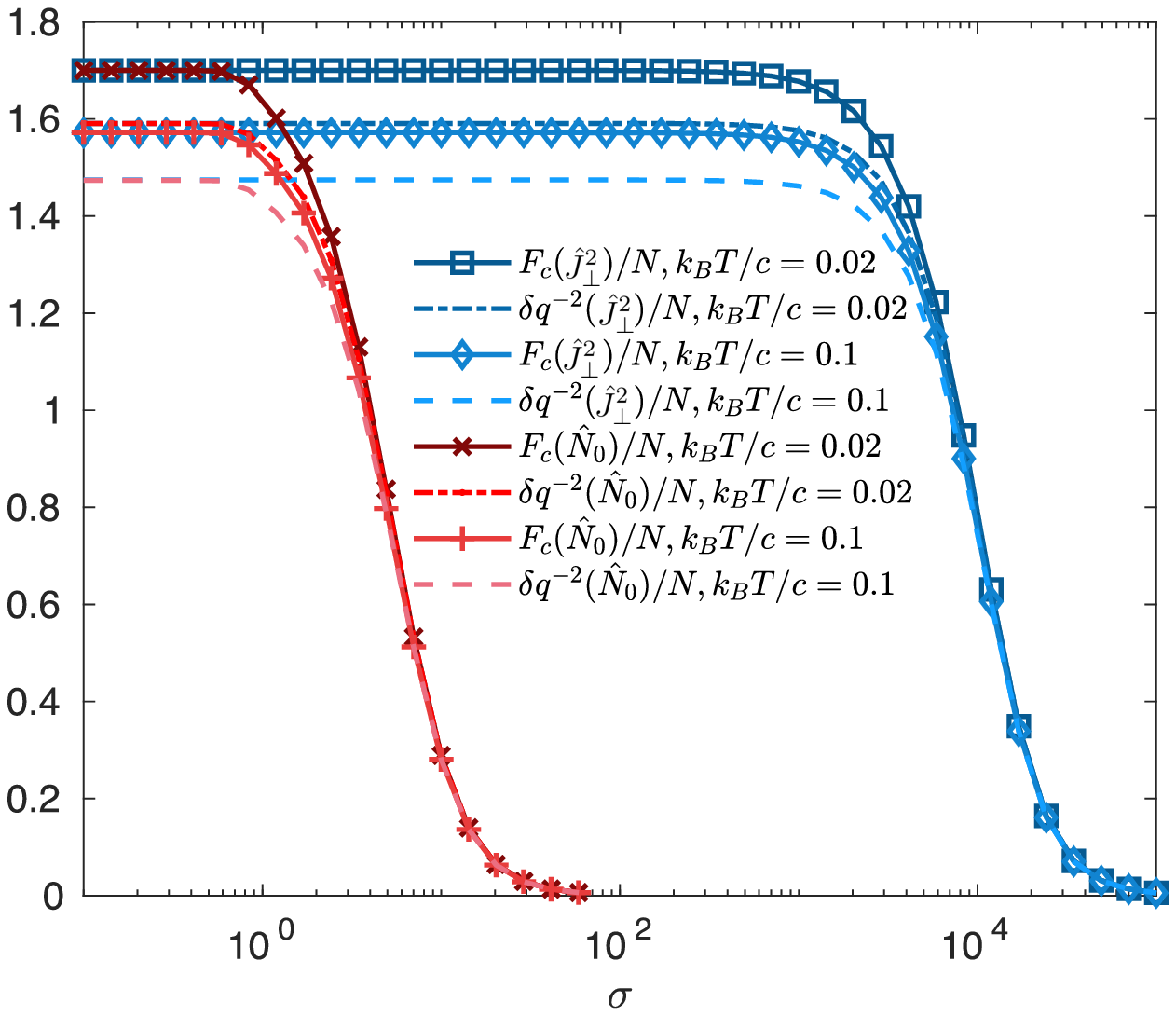}}
	\put(90,387){(a)}
	\put(-130,-8){\includegraphics[width=.52\textwidth]{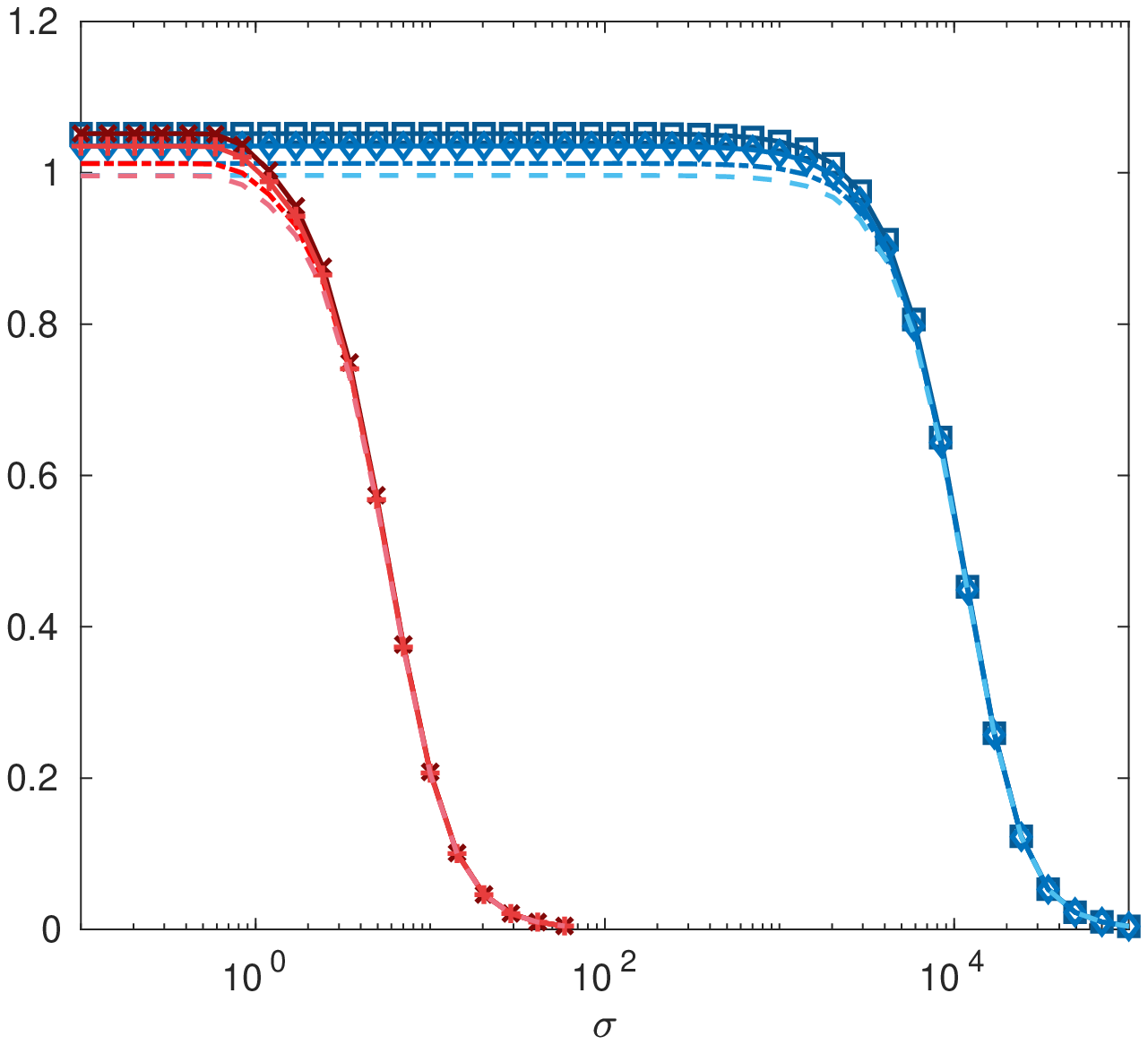}}
	\put(90,180){(b)}
	\end{picture}
	\caption{Effect of detection noise on the best sensitivity at $k_BT/c=0.02$ and $k_BT/c=0.1$ around the left (a) and right (b) critical points. The maximum values of the QFI, CFI and $\delta q^{-2}$ versus detection noise $\sigma$ are given when $\hat{\mathcal{S}}=\JPerpSq$  and $\hat{N}_0$.
	Here, the total number of atoms is $N=500$ and for the sake of simplicity we have removed the superscript $\max$ in the legend.}
	\label{fig:fig5bc}
\end{figure}

In Fig.~\ref{fig:fig5bc}, we show how the maxima of the CFI and $\delta q^{-2} $ varies with the strength of the noise $\sigma$ for both signals $\hat{\mathcal{S}}=\hat{N}_0$ and $\hat{\mathcal{S}}=\JPerpSq$, when $N=500$ and $k_BT/c=0.02$ (quantum regime) and $k_BT/c=0.5$ (classical regime). 
The sensitivity decreases by increasing the detection noise, as expected. 
In the limit of totally imperfect detectors, ($\sigma\gg 1$) the sub-SQL enhancement is lost. 
In this case, we see that the detection noise dominates the temperature effect and different curves for a fixed signal $\hat{\mathcal{S}}$ overlap. However, for both of the left (a) and right (b) critical points, the depth of change strongly depends on the chosen signal. Assuming a same value of $\sigma$ in both scenarios, this could imply that the $\hat{\mathcal{S}}=\JPerpSq$ is more robust against the presence of detection noise than $\hat{N}_0$. However, this is a {false analogy} {due to different nature of signals}. In the next subsection we discuss this point in more details.

\subsection{Analysis of the noise effect}\label{sec5C}

From an experimental point of view, our results suggest that one needs to achieve both a sufficiently cold initial sample, and a low detection noise in order to get the sub-SQL sensitivity. In the following, we discuss these limits and their implications in real experiments.

Regarding the temperature of the sample, thermometry in trapped bosonic quantum gases well below the condensation point is a difficult task. This occurs when the fraction of thermal atoms is negligible and therefore impossible to be distinguished from the condensate part. Some record low temperature measurements have been achieved by different techniques \cite{kurnthermometry,leanhardt2003cooling} relying on the thermal fraction of the system components. However, a reliable and simple experimental way of measuring temperatures of trapped Bose-Einstein condensates is still missing. Despite of that, the proposed protocol gives rise to collisional-induced entanglement \cite{luo2017deterministic}, which is easier to detect and is a clear indication that the phase transition is crossed adiabatically implying that $ k_B T / c < \Delta_{\rm min}$~\cite{Gerbier2016}.

\begin{figure*}[htb!]
	\begin{picture}(0,160)
	\put(-266,0){\includegraphics[width=.37\textwidth]{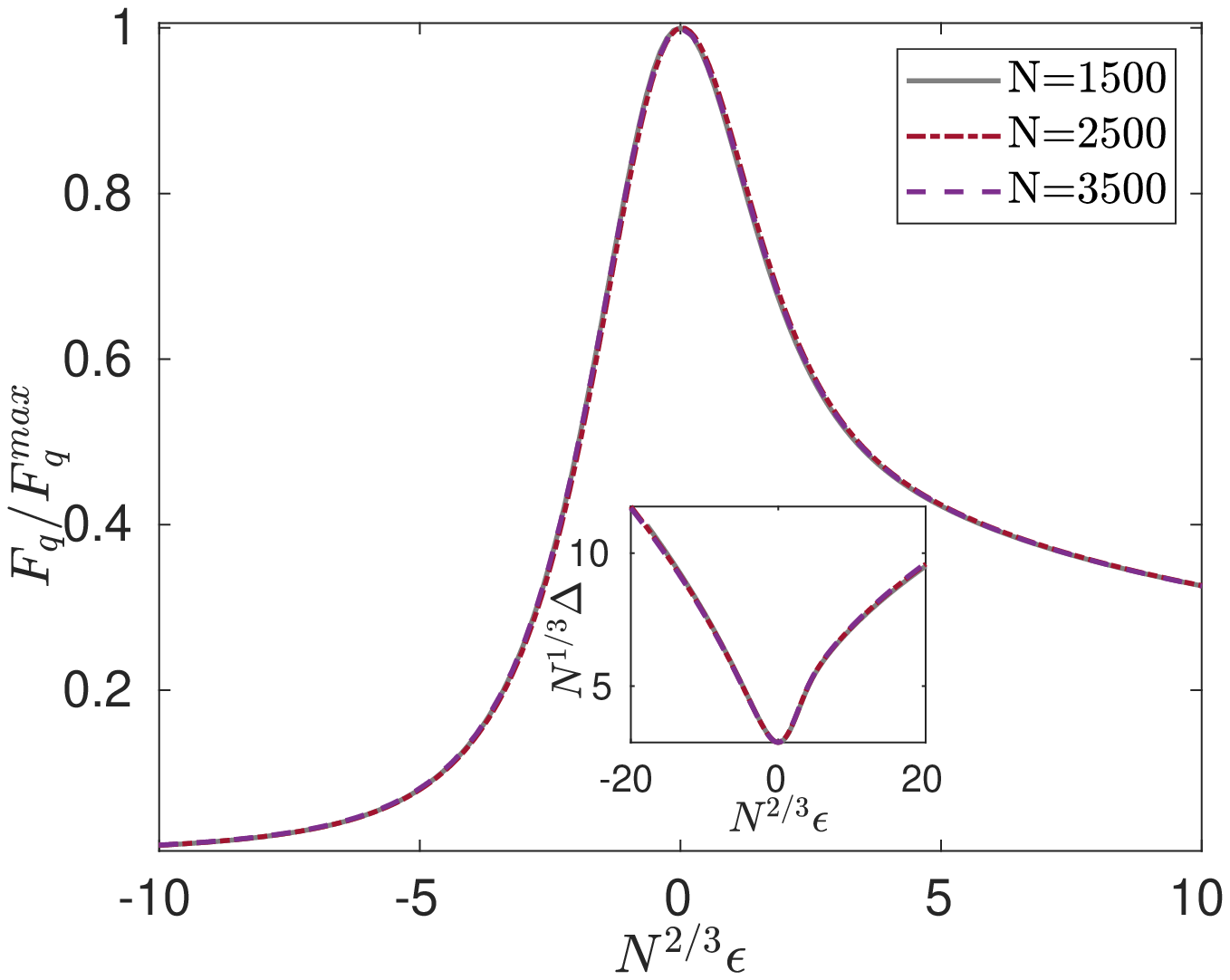}}
	\put(-236,127){(a)}
	\put(-93,0){\includegraphics[width=.37\textwidth]{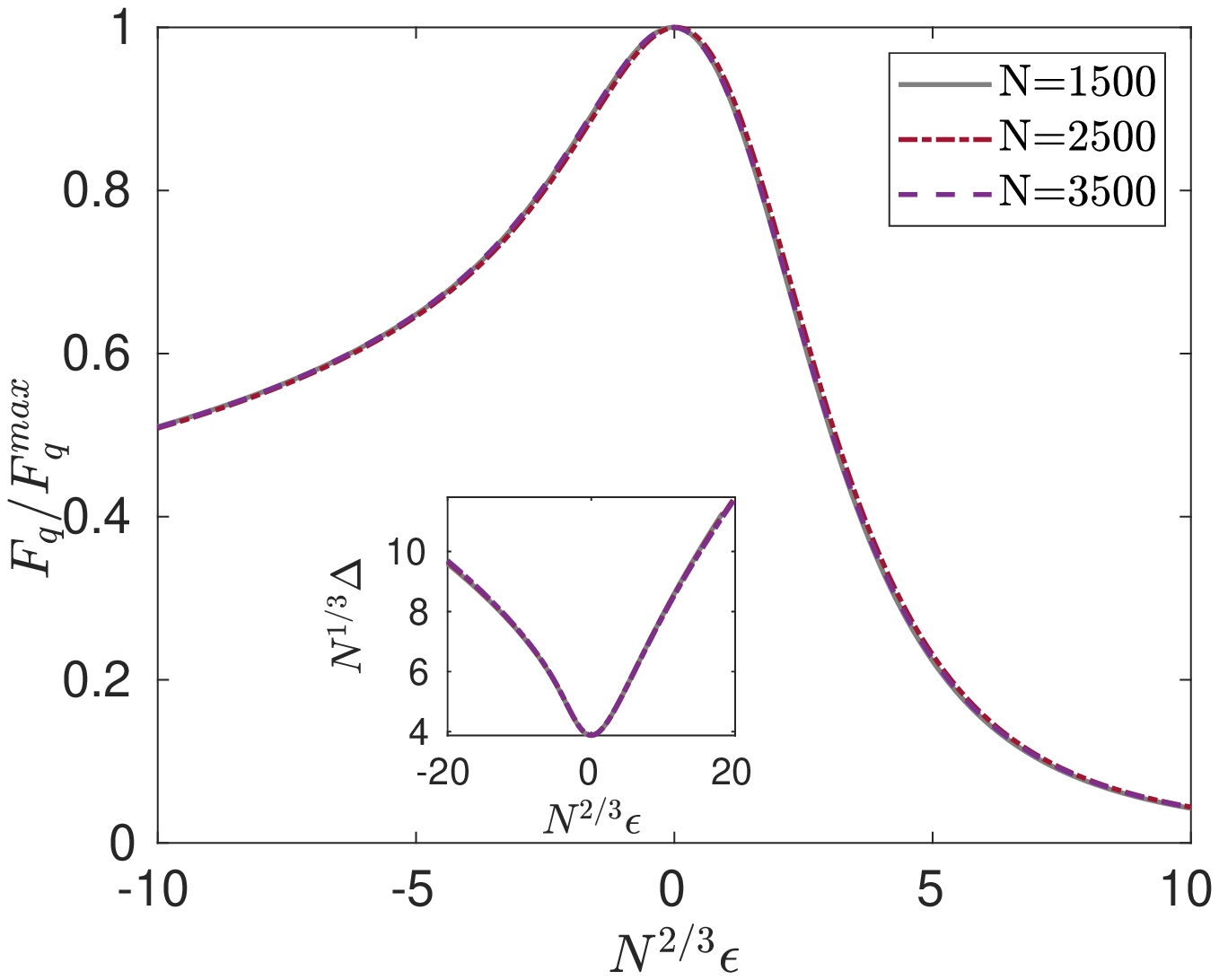}}
	\put(-63,127){(b)}
	\put(80,0){\includegraphics[width=.37\textwidth]{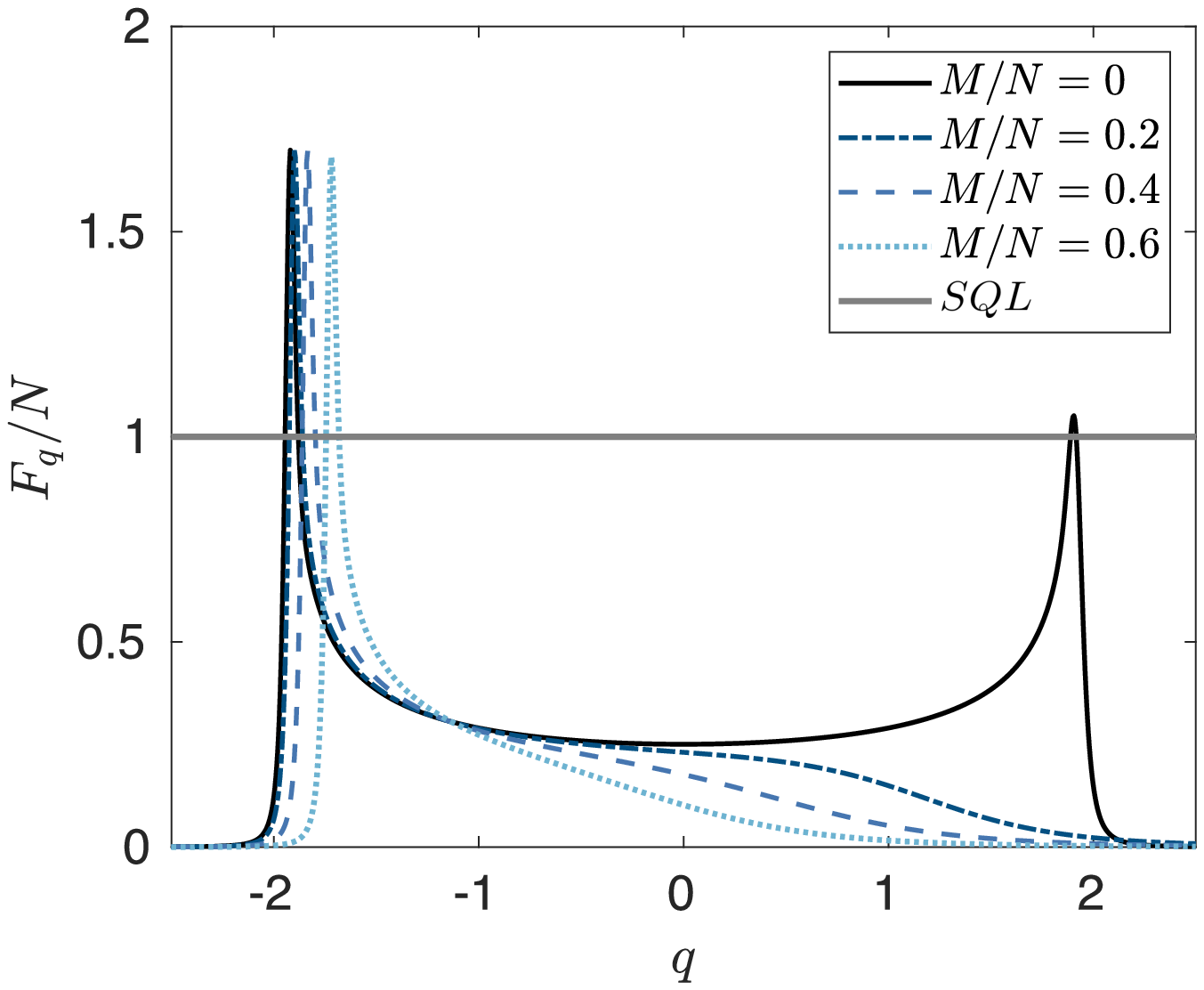}}
	\put(110,127){(c)}
	\end{picture}
	\caption{
		Scaling of the QFI around the left (a) and right (b) critical points for different number of atoms as indicated in the legend. We observe that $F_q = F_q^{\rm max} g(N^{\gamma}\epsilon )$, where $g(N^{\gamma}\epsilon )$ is an universal function and $\epsilon=q-q\supmax$ is a distance from the critical point. The results confirm $F_q^{\rm max} \sim N^{1.33}$ and $\gamma = 2/3$. 
		Insets demonstrate scaling of the energy gap $\Delta_{\rm min}$. 
		(c) The QFI for various magnetizations as indicated in the legend. The right maximum of the QFI disappears because there is no phase transition around $q=2$ for macroscopic magnetization, see in Fig.\ref{fig:fig1}.}
	\label{fig:fig6}
\end{figure*}

Additionally, 
Fig.~\ref{fig:fig5bc} suggests that an experiment based on atom number counting i.e measuring $\hat{N}_0$ would require the detection noise $\Delta N \leq 6$ atoms. Single atom imaging resolution has been achieved in the context of single trapped atoms and optical lattices using fluorescence imaging \cite{singleA,sherson2010single}, and also in the context of mesoscopic ensembles in a cavity, where the number of atoms is determined from shifts in the cavity frequency \cite{zhang2012collective}. More recently, near single atom resolution has been achieved in trapped quantum gases \cite{qu2020probing}. Nevertheless, this measurement resolution requires very careful calibrations, post-processing of images in order to filter the background noise, and at best it is restricted to very low atom numbers (up to $\simeq 1000$). 

On the contrary, measurements of $\JPerpSq$ would require less demanding experimental conditions, and they have been demonstrated in ultra-cold quantum gases experiments using non-demolition paramagnetic Faraday rotation \cite{gajdacz2016preparation,bason2018measurement,palacios2018multi}. In this text, we propose to examine the technique used in \cite{palacios2018multi,gomez2019interferometric,gomez2020bose}. This type of measurement is remarkably different from the fluorescence imaging method, since our observable does not belong to the system under study, but it is coupled to it. In Faraday probing, the observable is the Stokes parameter $\hat{S}_y$ of the probe laser beam, that changes due to an induced bi-refringence effect caused by the atomic ensemble. This change corresponds to the rotation of the linear polarization, and it is proportional to the projection of the collective spin of the ensemble $\hat{J}$ along the propagation direction of the beam (see Appendix \ref{app:Faraday} for a detailed analysis). Measuring along one of the perpendicular directions, for instance $\hat{y}$, will allow to measure the amplitude of $\hat{J}_{\perp}$. Under the appropriate experimental conditions of detection and input state of the probe beam, the detection noise can be expressed as $\Delta \hat{J}_{\perp} = \sqrt{{2}/{G_1^2 N_L}}$, where $G_1$ is an experimental calibration factor that depends on the geometry of the probing beam, $N_L$ is the number of photons in the pulse, and $\sqrt{{2}/{N_L}}$ is the readout shot noise in the photodetector. For recent works in the literature, using about $10^6$ photons, and calibrated coupling factor $G_1 \approx 5 \times 10^{-7}$, the reported uncertainties are $\approx 10^3$. Although we measure $\hat{J}_{\perp}$, we are considering $\JPerpSq$ as our theoretical observable.   The variance of a function of a random variable can be approximated as $\Delta^2\JPerpSq \approx (2 \hat{J}_{\perp})^2 \Delta^2\hat{J}_{\perp}$. Now, we have to consider that this quantity, in our experimental realization will be zero for $|q| > 2$, and at the two critical points $q \sim 2,-2 \rightarrow |\hat{J}_{\perp}| \simeq \mathcal{O}(1)$ as shown in Fig.~\ref{fig:fig2}. Therefore, at the phase transition $\Delta^2 (\JPerpSq) \sim \Delta^2 \hat{J}_{\perp}$. According to Fig.~\ref{fig:fig5bc}, detection noise at this level $\sigma \simeq 10^3$, would result in negligible effect in the measurements of CFI and $\delta q^{-2}$. 

All in all, achievable detection noise limits still allow the successful realization of such an experiment using both observables. However, in the case of measuring $\hat{N_0}$, it requires a lot of effort to achieve such a resolution, and increasing the total number of atoms in the sample beyond $\sim 10^3$ will make it even more complex. On the other hand measuring $\JPerpSq$ using Faraday imaging, would allow for a more robust scheme, which in addition, can support samples of $\sim 10^5 - 10^6$ atoms. Another possible advantage of Faraday imaging is its non-destructive nature, that allows to probe the system continuously, measuring within the same experimental realization the continuous change in the observable as a function of the control parameter \cite{colangelo2017simultaneous}.

\section{Summary and Conclusion}\label{sec6}

In this work, we showed how the sensitivity in measuring the control parameter $q$ can be enhanced around critical points in the ferromagnetic spin-1 Bose-Einstein condensate. In order to quantify the sensitivity in the estimation of the coupling constant, we used the quantum and classical Fisher information and the error propagation formula. We explored two different measurement observables, namely the atomic population in $\ket{F=1,m_f=0}$ Zeeman substate and the total spin $\JPerpSq$ when the longitudinal magnetization is fixed to zero.

We paid special attention to the scaling properties of sensitivity close to the critical points, where the maximum value of the QFI is expected to scale with the total atom number as $F_q^{\rm max}\sim N^\mu$. Our results confirm this property by numerically extracting the maxima of the QFI and obtaining $F_q^{\rm max} \sim N^{4/3}$. 
Away from criticality, however, we recover the classical SQL scaling $F_q^{\rm max} \sim N$. 
It is worth noting that the overall variation of the QFI versus $q$ is subject to a scaling law ${F_q}/{F_q^{\rm max}} = g(N^{\gamma}\epsilon )$, where $g(x)$ is the scaling function, $\epsilon=q-q\submax$ is a distance from critical point and $\gamma$ is a scaling exponent~\cite{cardy1996,Kwok2008}. This is demonstrated in Fig. \ref{fig:fig6} using $\gamma=2/3$.  Similar value of scaling exponent is found for other systems belonging to the same universality class,~e.~g.~Lipkin-Meshkow-Glick~\cite{Kwok2008}, Dicke~\cite{Liu2009}, Bosonic Josephson junction~\cite{Pezze2019} and Hamiltonian of the antiferromagnetic condensate~\cite{safoura2020} around second-order phase transitions. All of these systems are \emph{fully-connected} models with no spatial degrees of freedom~\cite{Duan2013}. {Our system has the same properties and this could suggests that our system belongs to the same universality class as the ones mentioned above.}
Moreover, the same scaling laws exponents are valid also for CFI and $\delta q^{-2}$~\cite{Pezze2019}, which we have proved analytically for $q\sim 0$ making use of perturbation theory.

Furthermore, we investigated the effects of temperature and detection noise on the sensitivity. In particular, we discussed that the effect of finite temperature gives rise to  different regimes of sensitivity depending on the value of the energy gap compared to the temperature. In the low temperature limit $k_BT/c\ll \Delta_{min}$ the sensitivity given by the QFI (and similarly the CFI and signal-to-noise ratio) is quite robust against thermal noise. By increasing the temperature, the sensitivity diminishes and eventually approaches the SQL. At finite temperature, we also noticed the appearance of a dip around $q\sim 0$ for $\hat{\mathcal{S}}=\JPerpSq$ which is explained analytically using perturbation theory. On the other hand, we have included the effect of detection noise and evaluated the sensitivity respective to the two different observables. Our results indicate that measurements of the total spin operator $\JPerpSq$ are more robust than measurements of the $\hat{N}_0$ population for samples with the same number of particles.
Let us remind that while we concentrate our work on the zero magnetization case, samples with finite magnetization can also be used. In this case, the underlying phase diagram restricts to the left critical point for the BA/AFM transition. This behaviour can be reflected in the behavior of the QFI for different macroscopic magnetization which is presented in Fig.~\ref{fig:fig6}. Note that the right peak only appears in the case of $M=0$, while the left peak moves rightward in equivalence to the phase diagram given in Fig. \ref{fig:fig1}.

Last but not least, our work suggests the feasibility of experiments with sensitivities below the SQL exploiting criticality. In this sense, a direct application could be the precise estimation of the critical point $q_c$ with sub-SQL sensitivity, or indirect evaluation of quantities that determine it. Moreover, the criticality of the system can lead to an atomic amplification process that boosts a weak signal which is not be detectable because of the noise \cite{Jacob2019,Jacob2019b,Moelmer2011}. In addition, this work can provide a way for potential applications in the context of quantum thermometry ~\cite{sanpera2018,sanpera2014,Moelmer2011} in spinor BEC systems.

\section{Acknowledgments}
We gratefully acknowledge fruitful discussions with Mohammad Mehboudi. This work was supported by Spanish MINECO projects OCARINA (Grant No. PGC2018-097056-B-I00), Q-CLOCKS (Grant No. PCI2018-092973), and the Severo Ochoa program (Grant No. SEV-2015-0522); 
Generalitat de Catalunya through the CERCA program; Ag\`{e}ncia de Gesti\'{o} d'Ajuts Universitaris i de Recerca Grant No. 2017-SGR-1354;  Secretaria d'Universitats i Recerca del Departament d'Empresa i Coneixement de la Generalitat de Catalunya, co-funded by the European Union Regional Development Fund within the ERDF Operational Program of Catalunya (project QuantumCat, ref. 001-P-001644); Fundaci\'{o} Privada Cellex; Fundaci\'{o} Mir-Puig; 17FUN03-USOQS, which has received funding from the EMPIR programme co-financed by the Participating States and from the European Union's Horizon 2020 research and innovation programme, Grant No. PCI2018-092973 (DBO), the Polish National Science
Center Grants DEC-2015/18/E/ST2/00760 (SSM) and QuantEra project MAQS, Grant No. UMO-
2019/32/Z/ST2/00016 (EW).

\appendix

\section{Analytical proof of equation (\ref{err}) for~$M=0$}\label{app:explanation}

In this appendix, we prove that when the system (\ref{H})
is in the ground state, the error propagation formula (\ref{delta}) for both $\hat{N}_0$ and $\JPerpSq$ signals leads to the same result (\ref{err}). 
Here, we consider the system in the subspace of zero magnetization ($\ave{\hat{J}_z}=0$ and $\ave{\hat{J}^2_z}=0$) which implies $\ave{\hat{J}^2_{\perp}}=\ave{\hat{J}^2}$.  

\begin{figure}[]
    \centering
	\includegraphics[width=1.0\columnwidth]{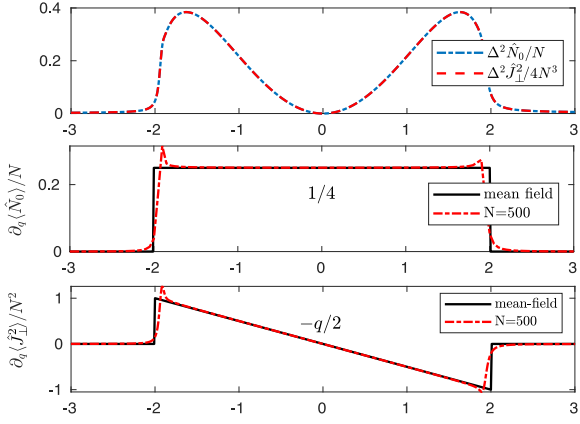}
	\caption{
	Numerical verification of various relations used in Appendix \ref{app:explanation}. (a) Numerical demonstration of (\ref{eq:rel}),  
	(b) the first derivative of $n_0$ obtained with exact calculations (red dashed lines) and the mean-field analysis (\ref{mf_sig_n0}),
	(c) the first derivative of $j^2$ from exact numerical simulations and compared to the analytical mean-field expression (\ref{mf_sig_j2}), all for $N=500$.
	} 
	\label{fig:fig7}
\end{figure}

Let us start with the variance of the Hamiltonian
\begin{equation}\label{H_var}
\Delta^2\hat{H}=\ave{\hat{H}^2}-\ave{\hat{H}}^2
\end{equation}
which is zero for the system in the ground state. It can be expressed as 
\begin{eqnarray}
    \frac{\Delta^2\hat{H}}{c^2}&
    =&\frac{\Delta^2 \JPerpSq}{(2N)^2}+ q^2 \Delta^2 \hat{N}_0 \nonumber \\
    &+& \frac{q}{2N}\left( \ave{\{ \JPerpSq, \hat{N}_0 \}}-2\ave{\JPerpSq} \ave{\hat{N}_0} \right).
\end{eqnarray}
The expression on the right hand site can be shown to be
\begin{equation}\label{bra}
    q\left( \ave{\{ \JPerpSq, \hat{N}_0 \}}-2\ave{\JPerpSq} \ave{\hat{N}_0} \right) = -\frac{1}{N} \Delta^2 \JPerpSq,
\end{equation}
when using $q\hat{N}_0=-\frac{\JPerpSq}{2N} - \frac{\hat{H}}{c}$, and $\ave{\JPerpSq \hat{H}}=\ave{\JPerpSq}\ave{ \hat{H}}$. The latter can be extracted using the identity operator, $\hat{I}=|\psi_0\rangle\langle \psi_0|+\sum_{\alpha\ne 0}|\psi_\alpha\rangle\langle \psi_\alpha|$, where
$|\psi_0\rangle$ and $|\psi_{\alpha\ne 0}\rangle$ refers to the ground and excited states respectively. 
Therefore one has
\begin{equation}
    \frac{\Delta^2\hat{H}}{c^2}=q^2 \Delta^2 \hat{N}_0 - \frac{\Delta^2 \JPerpSq}{(2N)^2},
\end{equation}
which in the limit $\Delta^2\hat{H} \to 0 $ reads
\begin{equation}\label{eq:rel}
    \frac{\Delta^2 \JPerpSq}{(2N)^2} = q^2 \Delta^2 \hat{N}_0
\end{equation}
in agreement with numerical results given in Fig. \ref{fig:fig7}.

On the other hand, the average values of the two signals $\hat{\mathcal{S}}=\hat{N}_0$ and $\JPerpSq$ in the denominator of (\ref{err}) can be approximated on the mean field level which is expected to be valid for large $N$. The mean-field approach can be performed by expressing annihilation and creation operators in (\ref{H}) according to $\hat{a}_{m_f}\to \sqrt{N_{m_f}} e^{i \theta_{m_f}}$, where $N_{m_f}$ is the number of atoms in the $m_f$ Zeeman component and $\theta_{m_f}$ is the phase. 
It was shown that $\theta_{1}+ \theta_{1} - 2 \theta_{1} =0 $~\cite{Zhang2003}. 
In this case, the operator $\JPerpSq = N + \hat{J}^2_z + \hat{N}_0(2N + 1 -2 \hat{N}_0)+ 2 (\hat{a}_0^\dagger{}^2\hat{a}_1\hat{a}_{-1}+\hat{a}_0^2\hat{a}_1^\dagger\hat{a}_{-1}^\dagger)$ transforms to $\JPerpSq \to N+N_0 + 2 N_0(\sqrt{N_1}+ \sqrt{N_{-1}})^2$ while $\hat{H}/(c N) \to \mathcal{F}$ with the following energy functional
\begin{equation}
    \mathcal{F}(n_0,q)=-2 n_0 (1-n_0) - q n_0 .
\end{equation}
To obtain the above form we have introduced fractional population of $m_f$-th Zeeman level $n_{m_f}=N_{m_f}/N$ and imposed the condition $M/N=n_1 - n_{-1}=0$, such that $n_1=n_{-1}$ and consequently $2n_1=1-n_0$ due to $n_1+n_0+n_{-1}=1$. 
Minimization of the energy functional with respect to $n_0$ approximates the mean value of the atoms number in the $m_F = 0$ Zeeman component. It gives 
\begin{equation}\label{mf_sig_n0}
n_0(q) = \left\{ \begin{array}{ll}
0, & \quad q<-2,\\
\frac{q}{4}+\frac{1}{2}, & \quad  q\in[-2,2] ,\\
1, & \quad q>2.
\end{array} \right. 
\end{equation}
On the other hand, the energy functional can be expressed in terms of the mean field value of $\ave{\JPerpSq/N^2} \to j^2$, which gives $j^2 = 4n_0(1-n_0)$ and sets $n_0=(1+\sqrt{1-j^2})/2$. Therefore, it is enough to use the latter in (\ref{mf_sig_n0}) to obtain variation of $j^2$ versus $q$. More formally the energy functional $ \mathcal{F}$ can also be expressed in terms of $j^2$ to casts in the following form 
\begin{equation}
    \mathcal{F}(j^2,q) = -\frac{j^2}{2} - \frac{q}{2} \left(1+\sqrt{1-j^2} \right).
\end{equation}
Minimization of the above expression with respect to $j^2$ leads to
\begin{equation}\label{mf_sig_j2}
j^2 = \left\{ \begin{array}{ll}
0, & \quad |q|>2,\\
1 - \frac{q^2}{4}, & \quad  q\in[-2,2] .
\end{array} \right. 
\end{equation}
In Fig.~\ref{fig:fig2} we compare the mean-field expressions (\ref{mf_sig_n0}) and (\ref{mf_sig_j2}) to exact quantum numerical results for $N=500,1500$. The excellent agreement can be noticed. 

All in all, the derivatives of $\ave{\hat{N}_0}/N$ and $\ave{\JPerpSq}/(N^2)$, which are present in the denominator of (\ref{err}), proved to fulfill the relation
\begin{equation}\label{eq:mfderiv}
\left|\frac{\partial_q \ave{\JPerpSq}}{2N}\right|^2 \approx \left| q \, \partial_q \ave{\hat{N}_0}\right|^2
\end{equation}
when the mean-field results for the average values of both signals, $\ave{\hat{N}_0}/N \approx n_0$ and $\ave{\JPerpSq}/(N^2) \approx j^2$, are used.
Consequently, the relation between signal-to-noise ratio for both signals considered here (\ref{err}) is proved taking into account (\ref{eq:mfderiv}) and (\ref{eq:rel}). 

The derivation we presented here is performed for the system ground state, but we can expect that it also holds in the low temperature limit at the canonical ensemble. 

\section{Sensitivity at $q\to 0$: perturbative approach}\label{app:perturbation}

In this appendix, we provide eigenstates and eigenvalues of the Hamiltonian (\ref{H}) around $q=0$ using perturbation theory. Next, we use them to analytically extract the sensitivity in the same vicinity. 

It is convenient to consider the BEC spinor system in the Dicke states basis which are equivalent to the eigenstates of total spin operator $\hat{J}^2| N, \mathcal{J}, M \rangle= \mathcal{J}(\mathcal{J} + 1) | N,  \mathcal{J}, M \rangle$, and its $z$-projection $\hat{J}_z| N, \mathcal{J}, M \rangle= M | N, \mathcal{J}, M \rangle$, with $\mathcal{J}\in [M,N]$. 
Equivalently, one could use the
Fock basis which are the eigenstates of atomic number operators $\hat{N}_{m_f}$, namely $|n\rangle= |N_{+1},N_0,N_{-1}\rangle = |n, N+M-2n, n-M\rangle$.
The Dicke states are defined as \cite{Niezgoda2019,Gerbier2012}
\begin{eqnarray}
|N,\mathcal{J},M\rangle=\frac{1}{\mathcal{N}^{1/2}}(\hat{J}_{-})^P(\hat{A}^\dagger)^Q(\hat{a}^\dagger_{+1})^\mathcal{J}|\text{vac}\rangle ,
\end{eqnarray}
where $P=\mathcal{J}-M$, $2Q=N-\mathcal{J}$ while $\hat{J}_{-}=\sqrt{2}(\hat{a}_{-1}^\dagger\hat{a}_0 + \hat{a}^\dagger_0 \hat{a}_1)$ is the spin lowering operator, $\hat{A}^\dagger=\hat{a}_0^\dagger-2\hat{a}_{-1}^\dagger \hat{a}_{+1}^\dagger$ is the singlet spin operator 
and $|\text{vac}\rangle$ refers to the vacuum Fock state. 
The respective normalization factor is given by
\begin{eqnarray}
\mathcal{N}=\frac{\mathcal{J}!(N-\mathcal{J})!!(N+\mathcal{J}+1)!!(\mathcal{J}-M)!(2\mathcal{J})!}{(2\mathcal{J}+1)!!(\mathcal{J}+M)!}
\end{eqnarray}
where $!!$ represent the double fractional. 

In order to analyze the sensitivity around $q=0$, we set $q$ as a small parameter of the perturbation theory. 
Consequently,
$\hat{H}_0=-\hat{J}^2/(2N)$ is the unperturbed Hamiltonian while $\hat{H}_q=-\hat{N}_0$ represents the perturbation. Then, the eigenstates of the unperturbed Hamiltonian can be considered as the Dicke states correspondent to  eigenvalues $E_\mathcal{J}=-\mathcal{J}(\mathcal{J}+1)/2N$. 
The lowest energy state is when $\mathcal{J}=N$ which can be expressed in terms of the Fock state basis as~\cite{Niezgoda2019} 
\begin{equation}\label{eq:DickinFock}
|N,\mathcal{J}=N,M=0\rangle=\sum_{n=0}^{N/2} c_n |n, N-2n,n \rangle,
\end{equation}
with
\begin{equation}
    c_n = \frac{2^{\frac{N}{2}-n}N!}{n! \sqrt{(N-2n)!}} \sqrt{\frac{N!}{(2N)!}}.
\end{equation}
Consequently, the eigenstates $| \psi_\mathcal{J} \rangle $ of the system Hamiltonian approximated by the perturbation theory up to the second-order correction of $q$ as 
\begin{align}
    |\psi_\mathcal{J}\rangle &= \left[ 1 - \frac{q^2}{2} 
    \left( C_-(\mathcal{J})^2 + C_+(\mathcal{J})^2 \right)  \right] | \mathcal{J} \rangle \nonumber \\
    &- q C_-(\mathcal{J})\left[ 1+ q F(\mathcal{J}) \right] | \mathcal{J}-2 \rangle \nonumber \\
    &- q C_+(\mathcal{J})\left[ 1+ q F(\mathcal{J}+2) \right] |\mathcal{J}+2 \rangle \nonumber \\
    &+ q^2 C_-(\mathcal{J}) C_-(\mathcal{J}-2) \frac{E_{\mathcal{J}-2}-E_{\mathcal{J}-4}}
    {E_{\mathcal{J}}-E_{\mathcal{J}-4}} 
    | \mathcal{J} - 4\rangle \nonumber \\
    &+ q^2 C_+(\mathcal{J}) C_+(\mathcal{J}+2) 
    \frac{E_{\mathcal{J}+2}-E_{\mathcal{J}+4}}
    {E_{\mathcal{J}}-E_{\mathcal{J}+4}} 
        | \mathcal{J} + 4\rangle,
        \label{eq:GS} 
\end{align}
for the case of $M=0$. Here, we introduced the notation $| \mathcal{J}\rangle = | N, \mathcal{J}, M=0 \rangle$.
Moreover, the respective eigenvalues $E_{\Psi_\mathcal{J}}$ read
\begin{align}
    &E_{\psi_\mathcal{J}} = E_\mathcal{J}
    - q \langle \mathcal{J}|\hat{N}_0|\mathcal{J} \rangle \nonumber \\
    &+ q^2 (E_\mathcal{J} - E_{\mathcal{J}-2})| C_-(\mathcal{J})|^2 +
    q^2 (E_\mathcal{J} - E_{\mathcal{J}+2}) |C_+(\mathcal{J})|^2. \label{eq:En}
\end{align}
In the above equations, we introduced the notations
\begin{align}
    C_-(\mathcal{J})&=\frac{\langle \mathcal{J}-2|\hat{N}_0 |\mathcal{J} \rangle}{E_\mathcal{J} - E_{\mathcal{J}-2}}, \nonumber \\
    C_+(\mathcal{J})&=\frac{\langle \mathcal{J}+2|\hat{N}_0 |\mathcal{J} \rangle}{E_\mathcal{J} - E_{\mathcal{J}+2}} ,\nonumber \\
    F(\mathcal{J})&=\frac{\langle \mathcal{J}|\hat{N}_0 |\mathcal{J} \rangle - \langle \mathcal{J}-2|\hat{N}_0 |\mathcal{J}-2 \rangle}{E_\mathcal{J} - E_{\mathcal{J}-2}}, \nonumber
\end{align}
in terms of
\begin{eqnarray}
\langle \mathcal{J}|\hat{N}_0|\mathcal{J}\rangle&=&A_+(\mathcal{J})+A_-(\mathcal{J}),  \label{N0} \\
\langle \mathcal{J}+2|\hat{N}_0|\mathcal{J}\rangle&=&\sqrt{A_{-}(\mathcal{J}+2)A_+(\mathcal{J})}, \\
\langle \mathcal{J}-2|\hat{N}_0|\mathcal{J}\rangle &=&\sqrt{A_+(\mathcal{J}-2)A_-(\mathcal{J})},
\end{eqnarray}
and
\begin{eqnarray}\nonumber
A_-(\mathcal{J})&=&\frac{(\mathcal{J}^2-M^2)(N+\mathcal{J}+1)}{(2\mathcal{J}-1)(2\mathcal{J}+1)},\\
A_+(\mathcal{J})&=&\frac{((\mathcal{J}+1)^2-M^2)(N-\mathcal{J})}{(2\mathcal{J}+1)(2\mathcal{J}+3)}.\nonumber
\end{eqnarray}
Note, that $\langle \mathcal{J}|\hat{N}_0|\mathcal{J}\pm 2\rangle=\langle \mathcal{J}\pm 2|\hat{N}_0|\mathcal{J}\rangle$~\cite{Niezgoda2019}.

In our calculations, we have also used the expression involving an average of~$\hat{N}_0^2$
\begin{align}
\langle \mathcal{J}|\hat{N}_0^2|\mathcal{J}\rangle 
&= [A_+(\mathcal{J})+A_-(\mathcal{J})]^2 \nonumber\\
&+A_-(\mathcal{J}+2)A_+(\mathcal{J})+A_-(\mathcal{J})A_+(\mathcal{J}-2).
\label{N02}
\end{align}

In the following we use these perturbative terms to derive the sensitivity of measurement $q$ around $0$.

\subsection{Zero temperature}

\begin{figure}[hbt!]
	\begin{picture}(0,110)
    \put(-125,0){\includegraphics[width=.26\textwidth]{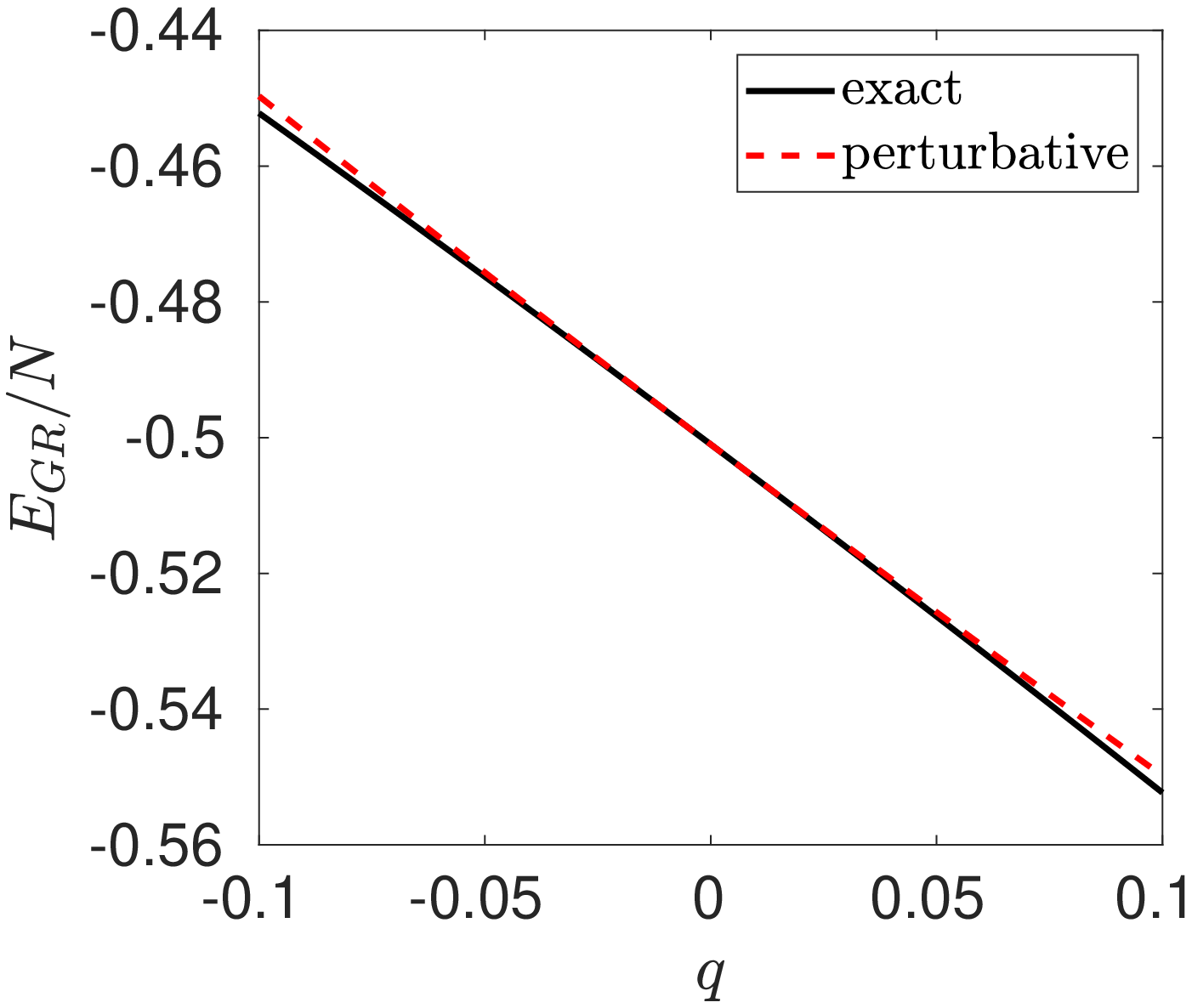}}
    \put(-90,90){(a)}
    \put(-1,0){\includegraphics[width=.27\textwidth]{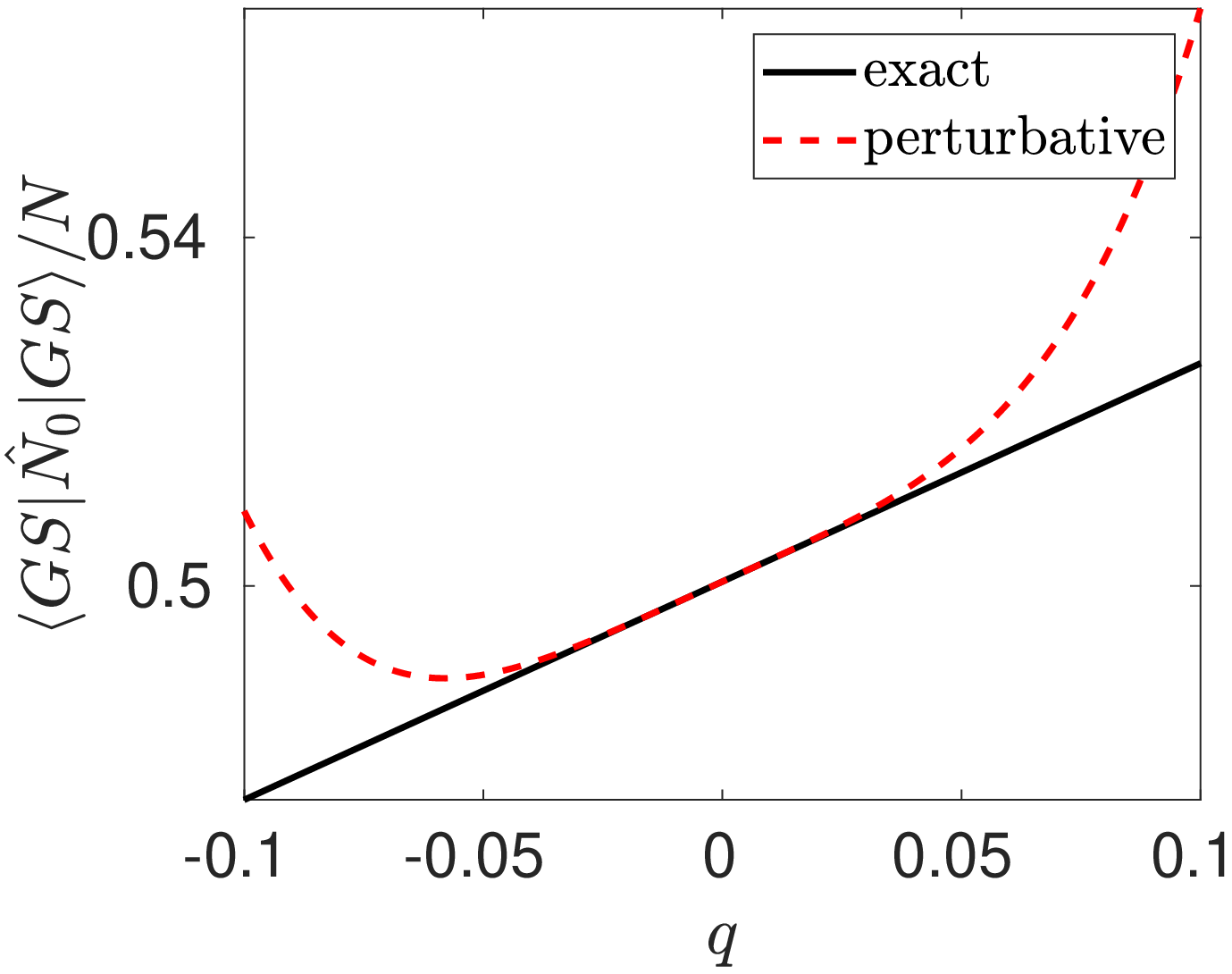}}
    \put(30,90){(b)}
    \end{picture}
	\caption{Validity of the perturbation theory for $N=500$. (a) The ground state energy. (b) An average value of $\hat{N}_0$ over GS. The exact numerical result is marked by the black solid line and analytical one from perturbation theory by the dashed red line.}
	\label{fig:fig8}
\end{figure}

In this subsection, we give the analytical results for the sensitivity around $q=0$ when the system is in the ground state, i.e. $T=0$. We employ perturbation theory and consider QFI, CFI and error propagation formula with $\hat{\mathcal{S}}=\hat{N}_0,\hat{J}^2$. In this case, the ground state energy of the system is for $\mathcal{J}=N$ (\ref{eq:GS}) and we explicitly get
\begin{equation}\label{E_p}
E_{\psi_N}= -\frac{N}{2}-q \frac{N}{2} - q^2\frac{N}{8}.
\end{equation}
Correspondingly, the ground state (\ref{eq:En}) is evaluated as 
\begin{equation}\label{psi_p}
|\psi_N\rangle = \left(1- q^2 \frac{N}{32} \right) |N\rangle 
+ q \frac{\sqrt{N}}{4}|N-2\rangle + q^2\frac{N}{16\sqrt{2}}|N-4\rangle ,
\end{equation}
in the large atoms number limit $(N \gg 1)$. 
In Fig.~\ref{fig:fig8} we demonstrate validity of the above approximated results for the ground state by comparing them to the exact numerical results for energy and an average value of $\hat{N}_0$.
Clearly, an agreement between the numerical and approximated results in the limit of $q \sim 0$ can be noticed.

In order to analyze the sensitivity around $q=0$, it is more straightforward to  consider an alternative expression for the QFI  valid for pure states \footnote{It can be derived from (\ref{QFI_fid}) using the Taylor expansion as shown in e.g.~\cite{Zanardi2007,Pezze:2016_review}.}
\begin{equation}\label{eq:zanardi}
    F_q=4 \left( \langle \partial_q \psi | \partial_q \psi \rangle - |\langle \psi | \partial_q \psi \rangle|^2 \right).
\end{equation}
Subsequently by replacing \ref{psi_p} and its derivative, one easily obtains
\begin{equation}\label{Fq_0}
    F_q |_{q\to 0}= \frac{N}{4}.
\end{equation}
On the other hand in order to get CFI, it is convenient to use the following definition
\begin{equation}\label{eq:classF}
    F_{c}(\hat{\mathcal{S}})=\sum_s \frac{1}{P(s|q)} \left( \frac{\partial P(s|q)}{\partial q}\right)^2.
\end{equation}
The CFI value depends on the particular choice of the operator $\hat{\mathcal{S}}$ as mentioned before. Let us start with $\hat{\mathcal{S}}=\hat{J}^2$. Using (\ref{psi_p}), probability $P(\mathcal{J}|q)=|\langle \mathcal{J}| \psi_N\rangle |^2$ reads
\begin{align}
P(\mathcal{J}|q)&=\left(1-q^2\frac{N}{32}\right)^2\delta_{\mathcal{J},N} \nonumber \\
&+ q^2 \frac{N}{4^2}\delta_{\mathcal{J},N-2} 
+ \left(\frac{q^2N}{16\sqrt{8}} \right)^2\delta_{\mathcal{J},N-4},\label{P_J}
\end{align}
where, $\delta_{\mathcal{J},\mathcal{J}'}$ refers to the Kronecker delta function. 
By inserting (\ref{P_J}) and its derivative with respect to $q$ into equation (\ref{eq:classF}), one gets 
\begin{equation}\label{FcJ2_0}
    F_{c}(\hat{J}^2)|_{q\to 0} = \frac{N}{4}.
\end{equation}
On the other hand, when the measurement signal is taken as the operator of number of atoms in the $m_F=0$ Zeeman component, $\hat{\mathcal{S}}=\hat{N}_0$, it is easier to work using the Fock state basis which is the eigenbasis of atomic number operators $\hat{N}_{m_F}=\hat{a}^{\dagger}_{m_F}\hat{a}_{m_F}$~\cite{Bigelow1998} as we show it in Eq.(\ref{eq:DickinFock}). 
Note, the coefficients of decomposition are normalized to one, i.e. $\sum_n |c_{n}|^2=1$.
The probability can be treated as
\begin{align}
&P(n|q)= |\ave{n|\psi_N}|^2 = \nonumber \\
&\left|(1-q^2 \frac{N}{32}) c_{n,N} + q\frac{\sqrt{N}}{4}c_{n,N-2} + q^2 \frac{N}{16\sqrt{2}} c_{n,N-4} \right|^2.\label{P_p}
\end{align}
After computing the derivative of $P(n|q)$ with respect to $q$, one can compute the CFI by
using (\ref{eq:classF}) and normalization condition for coefficients $c_n$. This leads to 
\begin{equation}\label{FcN0_0}
F_c(\hat{N}_0)|_{q\to 0} = \sum_n \frac{1}{c_{n,N}^2} \left( \frac{\sqrt{N}}{2}c_{n,N} c_{n,N-2} \right)^2 = \frac{N}{4}.
\end{equation} 

Consequently, putting all results together (\ref{Fq_0}), (\ref{FcN0_0}), and (\ref{FcJ2_0}), we get $F_Q=F_c(\hat{N}_0)=F_c(\hat{J}^2)$ when $q \to 0$.
It is in an agreement with the numerical results presented in Fig. \ref{fig:fig3}. Note, while the sensitivity around $q\to 0$ is of the order of SQL $\sim N$, the scaling around critical points exceed this limit, i.e. $\sim N^{4/3}$. This is due to the fact that the neighbour states differ significantly when varying $q$ around critical points, while they do not change much around $q\sim 0$. 

Finally, for the sake of completeness, we also derive the sensitivity considering error propagation formula (\ref{delta}).
Let us start with $\hat{\mathcal{S}}=\hat{J}^2$.
Using the perturbative states (\ref{psi_p}), we calculate the first $\langle \psi_N| \hat{J}^2 | \psi_N \rangle$ and the second $\langle \psi_N| \hat{J}^4 | \psi_N \rangle$ moments of the signal $\hat{\mathcal{S}}=\hat{J}^2$. By keeping the leading terms in $q$, we obtain the variance as $\Delta^2 \hat{J}^2 = q^2 N^3$ and the signal derivative as $|\partial_q \ave{ \hat{J}^2 }|^2= q^2 N^4/4$. The two latter result in $\delta q^{-2}=N/4$ for signal-to-noise ratio. 
In the case of the operator $\hat{\mathcal{S}}=\hat{N}_0$, the calculations can also be performed using (\ref{N0}) and (\ref{N02}) to obtain the first and second moments. The calculations are quite tedious while finally one finds that the leading terms in $q$ are the same as previously, namely $\Delta^2\hat{N}_0 = q^2 N^3$ and $|\partial_q \ave{\hat{N}_0}|^2= q^2 N^4/4$, which gives $\delta q^{-2}=N/4$.

\subsection{Finite temperature}\label{app:finiteT}

\begin{figure}[]
	\includegraphics[width=1\columnwidth]{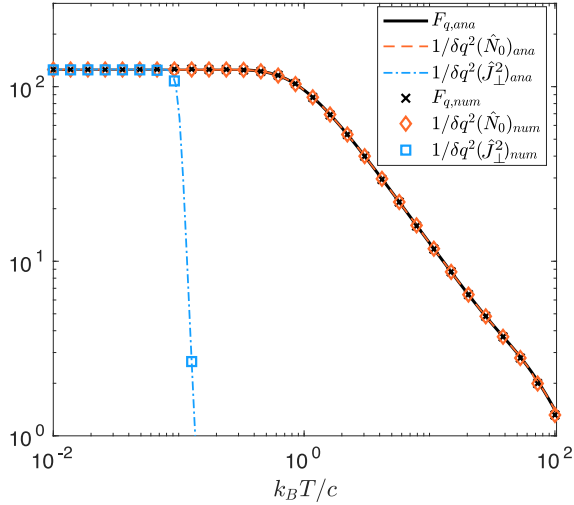}
	\caption{The sensitivity around $q\sim0$ versus temperature for $N=500$. 
	The analytical results for $F_{q,ana}$ (solid black), $\delta q^{-2}{(\hat{N}_0)}_{ana}$ (dashed red) and $\delta q^{-2}{(\JPerpSq)}_{ana}$ (dash-dotted blue) are compared to the exact results for $F_{q,num}$ (black crosses), $\delta q^{-2}(\hat{N}_0)_{num}$ (red diamonds) and $\delta q^{-2}(\JPerpSq)_{num}$ (blue squares). The abrupt drop of the sensitivity is observed when $\hat{\mathcal{S}}=\JPerpSq$ as explained in the text. Note, the log-log scale.  }
	\label{fig:fig9}
\end{figure}

Here, we explain the appearance of the dip around $q=0$ for $F_c(\JPerpSq)$ and $\delta q^{-2}(\JPerpSq)$ in the case of finite temperature. In the Dicke basis, a thermal equilibrium state of the system described by the density matrix (\ref{rho}) is given by
\begin{equation}\label{eq:rhoperturb}
\hat{\rho}=\sum_{\mathcal{J}} \omega_\mathcal{J} |\psi_\mathcal{J}\rangle \langle \psi_\mathcal{J}|,
\end{equation}
where, $\omega_\mathcal{J}=e^{-E_{\Psi_\mathcal{J}}(q)c/k_BT}/Z$. We take the high temperature limit, say $k_B T/c \to \infty$.

In finite temperature, it is useful to employ the following definition of the QFI~\cite{caves1994,Pezze:2016_review,Paris2016}
\begin{align}
    F_q&= 
    \sum_\mathcal{J} \frac{(\partial_q \omega_\mathcal{J})^2}{\omega_\mathcal{J}} 
    \nonumber \\
    {}&+2 \sum _{{\mathcal{J}},{\mathcal{J}'}\atop \omega_\mathcal{J}+\omega_{\mathcal{J}'}>0} \frac{(\omega_{\mathcal{J}} - \omega_{\mathcal{J}'})^2}
    {\omega_{\mathcal{J}} + \omega_{\mathcal{J}'}} 
    |\langle \psi_{\mathcal{J}'}| \partial_q \psi_{\mathcal{J}} \rangle|^2,
\end{align}
which is valid for mixed states. Making use of the above definition, we obtain 
\begin{align}
    &F_q|_{q \to 0} =
    \sum_\mathcal{J}  \frac{\tilde{\omega}_{\mathcal{J}}\, c}{k_B T}
    \left(\langle \mathcal{J}|\hat{N}_0 |\mathcal{J}\rangle - 
    \sum_\mathcal{J'}\langle \mathcal{J}'|\hat{N}_0 |\mathcal{J}'\rangle \tilde{\omega}_{\mathcal{J}'}  
    \right)^2 \nonumber \\
    & + 2 \sum_{\mathcal{J}} \left[
    \frac{(\tilde{\omega}_\mathcal{J} - \tilde{\omega}_{\mathcal{J} - 2})^2}
    {\tilde{\omega}_\mathcal{J} + \tilde{\omega}_{\mathcal{J}-2}} C_-^2(\mathcal{J}) +
    \frac{(\tilde{\omega}_\mathcal{J} - \tilde{\omega}_{\mathcal{J} + 2})^2}
    {\tilde{\omega}_\mathcal{J} + \tilde{\omega}_{\mathcal{J}+2}} C_+^2(\mathcal{J}) \right]\label{rho_JJ}
\end{align}
where, $\tilde{\omega}_{\mathcal{J}}:=\omega_{\mathcal{J}}|_{q\to 0}$.  
The above expression shows that $F_q$ is finite around $q=0$ when the temperature is non-zero, even if it is very small. This is because the coefficient in the first line is finite since both of $\tilde{\omega}_{\mathcal{J}}$ and $T$ are non-zero. while, the expression inside parenthesis is non zero because $\mathcal{J}\neq\mathcal{J}'$ and $\tilde{\omega}_{\mathcal{J}}<1$. On the other hand, one can observe that the QFI tends to zero in the high temperature limit, i.e $k_B T/c\to \infty$. This is due to $\partial_q \tilde{\omega}_\mathcal{J}|_{q\to 0 \atop T\to \infty} = 0$ 
while $\tilde{\omega}_{\mathcal{J}} - \tilde{\omega}_{\mathcal{J}'}|_{q\to 0 \atop T\to \infty} \ \rightarrow 0$ as a result of of having a totally mixed state. In Fig.~\ref{fig:fig9} we present the analytical and numerical result for the QFI value around $q=0$ versus the temperature value. The perfect agreement can be noticed.

Moreover, in order to get the sensitivity when $\hat{\mathcal{S}}=\JPerpSq$ ($q=0$), we use the signal-to-noise ratio (\ref{delta}). We start with the variance $\Delta^2 \hat{J}^2=\ave{\hat{J}^4} - \ave{\hat{J}^2}^2$ with $\ave{\hat{J}^{2l}}={\rm tr}(\hat{\rho}\hat{J}^{2l})$ ($l=1,2$). Using (\ref{rho_JJ})
, it is quite easy to show that the variance reads
\begin{equation}
    \Delta^2 \hat{J}^2 |_{q\to 0} 
    = \sum_\mathcal{J} \tilde{\omega}_{\mathcal{J}} \mathcal{J}^2 (\mathcal{J} + 1)^2 
    - \left[ \sum_\mathcal{J} \tilde{\omega}_{\mathcal{J}} \mathcal{J} (\mathcal{J} + 1), \right]^2,
\end{equation}
which tends to zero for the pure state when $\mathcal{J}=N$ (see previous subsection). However any mixed state makes the variance non-zero, although it can be very small. One can also show that the derivative of an average value of $\hat{J}$ with respect to $q$ is equal to
\begin{equation}
    \partial_q \ave{\hat{J}}|_{q\rightarrow 0} = \sum_\mathcal{J}  \frac{\tilde{\omega}_{\mathcal{J}}\, c}{k_B T}
    \left(\langle \mathcal{J}|\hat{N}_0 |\mathcal{J}\rangle - 
    \sum_\mathcal{J'}\langle \mathcal{J}'|\hat{N}_0 |\mathcal{J}'\rangle \tilde{\omega}_{\mathcal{J}'}  
    \right).
\end{equation}
Consequently, for any non-zero temperature we get $\partial_q \ave{\hat{J}^2}|_{q\rightarrow 0} \rightarrow 0$  due to $\sum_{\mathcal{J}}\tilde{\omega}_{\mathcal{J}}=1$.
Therefore, the inverse of signal-to-noise ratio (\ref{delta}) gives
\begin{equation}
\delta q^{-2}|_{q\to 0}
=\frac{|\partial_q\ave{\hat{J}^2}|^2}{\Delta^2{\hat{J}^2}} |_{q\to 0}\ \rightarrow 0.
\end{equation}

In addition, we derive the CFI (\ref{eq:classF}) for $\JPerpSq$ in the vicinity of $q\sim 0$. Using equations (\ref{psi_p}), (\ref{E_p}) and (\ref{eq:rhoperturb}), one finds the following probability distribution  
\begin{eqnarray}\nonumber
    P(\mathcal{J}|q)&=&\langle \mathcal{J}|\hat{\rho}|\mathcal{J} \rangle\\ \nonumber
    &=&\sum_{\mathcal{J}} \omega_\mathcal{J}
    \left[ 1 - \frac{q^2}{2} \left( C_-^2(\mathcal{J}) + C_+^2(\mathcal{J}) \right)\right],\\
\end{eqnarray}
which $P(\mathcal{J}, q)|_{q\to 0 } = 1$ and the respective derivative  as $\partial_q P(\mathcal{J}| q)|_{q\to 0} = 0$. Therefore, the above analysis shows that the CFI reads 
\begin{equation}
    F_{c}(\hat{J}^2)|_{q\to 0} \rightarrow 0,
\end{equation}
for any non-zero temperature.

Lastly, we show that the inverse of signal-to-noise ratio and the CFI for $\hat{N}_0$ gives a non-zero value. The first one can be shown to be nonzero, because the derivative in the perturbation theory reads
\begin{align}
    |\partial_q \langle \psi_N|N_0|  \psi_N  \rangle|_{q\to 0} 
    &= 2 \sum_{\mathcal{J}} \tilde{\omega}_\mathcal{J} 
    \left[ 
    C_+(\mathcal{J}) (E_\mathcal{J} - E_{\mathcal{J}+2})
    \right.\nonumber \\
    &\left. +
    C_-(\mathcal{J}) (E_\mathcal{J} - E_{\mathcal{J}-2})
    )\right]
\end{align}
which is a non-zero value since $E_{\mathcal{J}}\neq E_{\mathcal{J}\pm 2}$. Therefore, the variance is $\Delta ^2 \hat{N}_0|_{q\to 0}=\langle\mathcal{J}|\hat{N}_0^2|\mathcal{J}\rangle - \langle \mathcal{J}|\hat{N}_0|\mathcal{J}\rangle^2\neq 0$ using (\ref{N0}) and (\ref{N02}). We have not brought the final expression here because they are lengthy but instead in Fig.\ref{fig:fig9}, we demonstrate $\delta q ^{-2}(\hat{N}_0)$ depending on temperature when $q\sim 0$. Clearly, at any temperature $\delta q^{-2}(\hat{N}_0)$ is equal to the QFI value.
Consequently due to the inequality relation (\ref{eq:inequalities}), we conclude that the CFI for $\hat{\mathcal{S}}=\hat{N}_0$ equals to the QFI as well, at $q=0$ and for any temperature.\\

\section{Mathematical description of Faraday measurements}\label{app:Faraday}

As usual, we describe the system using the collective spin operators $\hat{J}_x, \, \hat{J}_y,\, \hat{J}_z$ and the light using the Stoke operators:
\begin{eqnarray}
\hat{S}_{\mu} = \frac{1}{2} (\hat{a}^\dagger_{+} \hat{a}^\dagger_{-}) \sigma_{\mu} (\hat{a}_{+} \hat{a}_{-})^T,
\end{eqnarray}
where $\sigma_{\mu}$ are the Pauli matrices, and $\hat{a}_{\pm}$ are anihilation operators of photons in $\sigma_{\pm}$ polarization states.
The interaction Hamiltonian describing atom-light interaction under dipole approximation can be decomposed into three parts proportional to the scalar, vector and tensor parts of the polarizability tensor $\hat{H}_{int} = \hat{H}_0 + \hat{H}_1 + \hat{H}_2$, see~\cite{phdKoschorreck, PhysRevA.73.042112} for mode details. The relevant interaction term is the vector part $\hat{H}_1$ (because $\hat{H}_0$ commutes with all Stokes parameters and the small magnitude of the $\hat{H}_2$ under certain experimental conditions i.e off resonant interaction and appropriate input polarization state):
\begin{equation}
\hat{H}_{1} = \frac{\hbar G_1}{\tau_p} \hat{S}_z \hat{J}_y,
\end{equation}
where $\tau_p$ is the probing time, and $G_1$ is a calibrated factor that takes into account the polarizability and the geometry of the beam, i.e it is like an effective coupling factor between the atoms and the probe beam. 
A pulse of off-resonant polarized light will experience a rotation of its polarization vector, compared to its quantum components.
The evolution of the Stokes operators under the given Hamiltonian can be calculated using the evolution operator $\hat{U}(t)= e^{-i \hat{H}_1 t/\hbar}$ following the usual prescription $\hat{S}_i(t) = \hat{U}^T \hat{S}_i(0) \hat{U}$.
When the input light is polarized along the $x$ axis, one has $\langle \hat{S}_y(0) \rangle = \langle \hat{S}_z(0)\rangle = 0$, $\langle \hat{S}_x(0) \rangle=N_L/2$ where $N_L$ the total number of photons at the input. 
The small change in the rotation angle $\phi$ can be defined as $\phi \approx \ave{\hat{S}_y}/\ave{\hat{S}_x}$. On the other hand, the evolution of $\hat{S}_y$ for a time $\tau_p$ can be written as:
\begin{equation}
\hat{S}_y(t) = \hat{S}_x (0) \sin G_1 \hat{J}_y .
\end{equation}

When one assume an initial atomic state fully polarized along the $z$-direction, perpendicular to the quantization axis $x$, then the following relation can be obtained
\begin{eqnarray}
\ave{\hat{J}_y} = \frac{1}{G_1} \frac{\ave{\hat{S}_y}}{\ave{\hat{S}_x}},
\end{eqnarray}
in the small angle approximation. This links the small rotation angle with an average value of atomic pseudo-spin component.

\bibliography{bibliography}

\end{document}